\documentclass{article}

\usepackage[preprint]{neurips_2026}

\usepackage[utf8]{inputenc} %
\usepackage[T1]{fontenc}    %
\usepackage{hyperref}       %
\usepackage{url}            %
\usepackage{booktabs}       %
\usepackage{amsfonts}       %
\usepackage{nicefrac}       %
\usepackage{microtype}      %
\usepackage{xcolor}         %
\usepackage{enumitem}

\usepackage{multirow}
\usepackage{booktabs}
\usepackage{array}
\renewcommand{\arraystretch}{1.3}
\usepackage{rotating}
\usepackage[table]{xcolor}
\usepackage{wrapfig} 
\usepackage{colortbl}
\usepackage{array}
\usepackage{hhline}
\usepackage{diagbox}
\usepackage{pifont}
\usepackage{natbib}
\usepackage{amsmath}
\usepackage{cleveref}

\Crefname{figure}{Fig.}{Figs.}
\crefname{figure}{Fig.}{Figs.}

\definecolor{lightblue}{RGB}{221,232,247}
\definecolor{miniaccentblue}{RGB}{200,222,244}
\definecolor{accentlightblue}{RGB}{173,200,233}
\definecolor{lightpink}{RGB}{249,233,247}
\definecolor{miniaccentpink}{RGB}{246,218,240}
\definecolor{accentlightpink}{RGB}{243,205,233}
\newcommand{\crossmark}{\ding{55}}

\title{Watermarks Attack Watermarks: Re-Watermarking as a Generic Removal Strategy}

\author{%
  Maria Bulychev, Neil G.\ Marchant, Benjamin I.\ P.\ Rubinstein \\
  University of Melbourne \\
  \texttt{\{m.bulychev, nmarchant, benjamin.rubinstein\} @unimelb.edu.au} \\
}

\begin{document}

\maketitle

\begin{abstract}
  Watermarking combines an imperceptible change to an input image that will trigger a detector, to assert provenance and protect intellectual property. The literature has shown great interest in attacks on watermarking schemes: attackers are clearly motivated to steal copyrighted material or circumvent legislated deepfake protections. In this work, we make a simple-yet-powerful observation: that such attacks on watermarking---like watermarks themselves---seek an imperceptible change to an input image (now already watermarked) that will trigger a detector. This analogy comparing watermark attacks to watermarking itself is highly suggestive: that watermarks could be used to attack watermarks. 
Our first contribution validates this hypothesis. 
In rigorous experiments\footnote{The project
is available at \url{https://github.com/MariaBulychev/Watermarks-Attack-Watermarks}} spanning 96 combinations of dataset, victim, and attack watermarks, %
we show that simply re-watermarking an already watermarked image reliably suppresses the original signal, without requiring gradients, surrogate models, or detection keys. 
Our second contribution is a simple classifier for detecting the presence and identity of an existing watermark in a given image. Surprisingly, experimental findings demonstrate outstanding overall accuracies 0.878--0.953.
This result is of independent interest as a security vulnerability: research shows that method-specific attacks achieve substantially stronger removal than black-box attacks.  
Taken together, watermark identification combined with re-watermarking successfully reduces bit accuracy by at least 25\% and up to 48\%. Our work
constitutes a cheap, generic, and highly effective attack pipeline, calling into question the reliability of current watermarking schemes to such a simple attack, as well as the value of existing sophisticated attacks.

\end{abstract}

\section{Introduction} \label{sec:intro}

Recent advances in generative AI have made it easy to synthesize photo‑realistic images and stylized artwork at scale, blurring the line between human‑created and machine‑generated media. Text‑to‑image systems such as Flux~\cite{blackforestlabs2025flux2}, Midjourney~\cite{midjourney2025v7}, and GPT-4o~\cite{openai2025gpt4oimg} can now produce an essentially unbounded variety of convincing visuals, amplifying long‑standing concerns around misinformation, deepfakes, intellectual‑property abuse, and the erosion of trust in digital content~\cite{guo_article}. In response, regulators have begun to mandate technical transparency: for example, the European Union's AI Act requires providers of generative systems to mark synthetic content so that it can later be identified as artificially generated or manipulated~\cite[Art. 50(2)]{eu-ai-act-2024}.

To meet these obligations, a growing ecosystem of invisible watermarking schemes has been proposed as a primary mechanism for asserting provenance and verifying authenticity %
in AI‑generated imagery~\cite{tree_rings, bui2023rosteals, sander2025watermark, soucek2025pixel,Tancik_2020_CVPR, zhang2024attackresilient}. Research has questioned the robustness of watermarks, by proposing a range of sophisticated attacks \cite{zhao2024invisible, saberi2024robustness, an2024waves}. We push this idea to its limit, by showing that watermarks can be cheaply erased by an attack that does nothing more than re‑watermark the image.

We make a simple yet powerful observation: watermark embedding and watermark removal solve nearly identical optimisation problems. Both seek a minimal perturbation within a perceptually bounded neighbourhood that steers a detector's decision---embedding pushes it toward detection, removal pushes it away. We hypothesize that watermark encoders are therefore, almost by construction, excellent tools for attacking other watermarks.
As watermarks are readily available and an attacker targeting watermarking would conceivably want to validate their attacks prior to deployment, we argue that access to watermarks is eminently reasonable to assume. Furthermore, our application of watermarks requires no specialized skills.
We perform rigorous experiments testing 8$\times$6 attack-victim pairs over 2 datasets and observe a clear pattern: post-processing watermarks can be removed by re-applying the method again on top of itself; in-processing watermarks are removed by applying the post-processing watermark ZoDiac \cite{zhang2024attackresilient} on top. Re-watermarking matches or outperforms state-of-the-art baseline attacks at significantly lower quality damage and additionally creating an ambiguity channel for ownership claims. We verify experimentally that the newly embedded message can still be recovered with high bit accuracy.

This “watermarks attack watermarks” phenomenon has two consequences that are especially troubling for provenance and copyright. First, re‑watermarking is an alarmingly cheap and generic removal strategy: an attacker does not need %
any sophisticated attacks, but only the ability to run a specific watermark encoder to overlay a new message. It is difficult to imagine an attacker with less capability than this; any adversary motivated to target watermarking systems can surely run a watermarking tool. As watermarks are made accessible to content creators to protect their IP, it is even conceivable that internet users at-large could undermine regulated watermarks.  Second, re‑watermarking does not only suppress the original signal; it also replaces it with a new, valid watermark message. As a result, whoever re‑watermarks the image last can plausibly claim authorship, while the original owner’s watermark is either undetectable or statistically dominated.

A natural objection is that a realistic attacker may not know which watermarking method was used on a given image. In our second contribution, we demonstrate that it is remarkably easy to train a classifier that, given a single image, identifies which watermarking algorithm produced it. We train what is, to our knowledge, the first classifier that identifies the specific watermarking method applied to an image from visual inspection alone, without access to the detection model, its parameters, or the embedding key. Achieving accuracies between 87.8\% and 95.3\%, it enables our attack pipeline to select the most effective overwriting strategy automatically. Together, these two contributions form a fully blind, end-to-end attack pipeline that reduces bit accuracy by at least 25\% and up to 48\% across all victim methods, while causing substantially less perceptual degradation than existing attacks.

The ability to distinguish between watermarking methods, beyond merely detecting their presence, is itself a vulnerability because it empowers adversaries with method-specific knowledge.
Comparing recent benchmarks like the NeurIPS 2024 “Erasing the Invisible”%
, we can observe that beige‑box attacks (which know the method but not its secret key or internal details) achieve stronger watermark removal than black‑box attacks \cite{shamshad2025firstplacesolutionneurips2024}. Additionally, recent work shows that knowledge of the watermarking method enables method-specific attacks that cause substantially larger drops in detection accuracy than generic attacks, even for schemes considered robust \cite{lin2025crack, shamshad2025firstplacesolutionneurips2024}.

Putting these pieces together, we argue that re-watermarking enables cheap overwriting of existing watermark signals. Combined with method identification, this allows effective removal or replacement of watermark evidence, calling into question the robustness of current approaches.

\section{Related Work}
Here, we focus on the work most directly related to our contributions; extended discussion of related work appears in \Cref{app:related}.

\paragraph{Watermark Coexistence} 
Petrov et al.~\cite{petrovcoexistence} 
show that classical and deep image watermarks can often be sequentially embedded with minor quality drops. However, they focus on ensembling watermarks to increase total capacity, rather than exploring adversarial overwriting.
In particular, they do not consider integrated watermarking for modern generative models such as latent diffusion or noise-space schemes, and they treat coexistence as a positive feature to be exploited for benign use cases rather than a potential vulnerability. Accordingly, their evaluation reflects a best-case perspective: watermarks are embedded cooperatively and assessed on how well they can coexist. By contrast, our work is the first to examine the worst case and to explicitly point out that (non‑)coexistence between watermarking methods can itself be exploited as an attack made practical by our watermark classifier.
\paragraph{Collision and Ambiguity Attacks} have been proposed as attack mechanisms in related domains. For LLM watermarks, \cite{luo2025lost} show that embedding multiple watermarks degrades detection accuracy and speculatively suggest repurposing a detector to drive such an attack, without pursuing it empirically. Ambiguity attacks~\cite{chen2023effective} instead target ownership claims by forging a competing watermark on top of the original, a different objective from erasure. Our work realizes the collision idea in the image domain: we confirm that watermarks interfere with one another and show that a classifier can identify which watermark is present to select the most damaging overwrite.

\paragraph{Watermark Detection and Undetectability} In work related to our contributed classifier, WMD~\cite{pan2025finding} is the first black‑box, annotation‑free detector for invisible image watermarks, distinguishing watermarked from non‑watermarked samples without knowledge of the underlying algorithm. WMD addresses \textit{whether} some watermark is present; our classifier instead identifies \textit{which} specific method was applied, then uses this to guide a targeted overwriting attack. Undetectability has been formalized cryptographically~\cite{christ2024undetectable}, and \cite{gunn2025undetectable} instantiate this guarantee for generative image models, showing empirically that all tested schemes leave detectable traces. Our classifier corroborates this: trained purely on visual features, it reliably identifies the applied watermarking method without any key or model access. Finally, in independent and concurrent work, \cite{senthuran2026srw} consider the self-re-watermarking threat model, demonstrating that existing methods consistently fail under same-encoder reapplication.
Our work differs in three key respects: we consider a broader attack surface covering both post-processing and in-processing watermarking schemes, we demonstrate that cross-method re-watermarking is also highly effective, and we show that a classifier can automate method identification to yield a fully blind end-to-end attack pipeline.

\paragraph{Black-Box and Beige-Box Attacks} The WAVES benchmark~\cite{an2024waves} established a comprehensive stress-testing protocol across 26 attack variants, finding regeneration-based attacks~\cite{zhao2024invisible, saberi2024robustness} among the most effective. 
However, method-specific beige-box attacks consistently surpass these generic strategies~\cite{shamshad2025firstplacesolutionneurips2024, serzhenko2025watermark}: knowledge of the embedding mechanism enables targeted interventions such as surrogate-detector attacks or phase perturbations that cause substantially larger drops in detection accuracy~\cite{lin2025crack}. The ability to identify the watermarking method is therefore a critical vulnerability---one our classifier directly provides.

\section{Preliminaries \& Threat Model}
\label{sec:threat_model}
In this section, we provide the necessary background on image watermarking, formalize the capabilities and goals of the adversary we consider, and establish the experimental evaluation framework shared across the subsequent sections.

\subsection{Image Watermarking}
From a system perspective, an invisible image watermarking scheme consists of an embedder and a detector. The embedder introduces a minimal, imperceptible perturbation to an input image, while the detector analyzes an image to determine if this specific perturbation is present \cite{hayes2018visible, wang2024adversarial}. Watermark designers face a fundamental trade-off: the embedded signal must be robust against common image distortions (like compression or cropping) while remaining formally and perceptually undetectable to human observers or automated analysis \cite{zhao2024invisible}.

Watermarking methodologies generally fall into two paradigms. \textit{In-processing} watermarks (e.g.,~Tree-Ring \cite{tree_rings}, Stable Signature \cite{Fernandez_2023_ICCV}) embed their signal during the generative process itself, directly modifying the latent space or generation trajectory of models like Stable Diffusion. Consequently, they can only be applied to AI-generated images at the time of creation. \textit{Post-processing} watermarks (e.g., StegaStamp \cite{Tancik_2020_CVPR}, RoSteALS \cite{bui2023rosteals}, ZoDiac \cite{zhang2024attackresilient}, Pixel Seal \cite{soucek2025pixel}) are model-agnostic and are applied to the image pixels or latents after generation, supporting both natural and AI-generated media. 
We introduce this distinction here because it constrains the attack surface: in-processing watermarks cannot be applied on top of an already generated image, limiting their use as an attack tool.

Furthermore, watermarking schemes differ in their capacity. Multi-bit schemes encode a recoverable cryptographic payload (e.g., 100 bits) linking the image to specific provenance metadata. By contrast, 0-bit schemes embed a fixed statistical trace that only allows the detector to emit a binary ``watermarked'' or ``clean'' decision. A detailed overview of the eight specific watermarking algorithms evaluated in this paper is provided in \Cref{app:method_details}.

\subsection{Threat Model}
Viewed structurally, watermark embedders and removal attacks solve closely related optimization problems: both seek a minimal, perceptually bounded perturbation that steers a detector's decision. We therefore hypothesize that applying a second watermark on top of an existing one constitutes a natural attack strategy. To formalize this, we define the \textit{victim} watermark as the original protection placed on the image, and the \textit{attack} watermark as the scheme used by the adversary to disrupt it.

\paragraph{Attacker Objectives} The primary goal of the attacker is \textit{erasure}: modifying the watermarked image such that the victim's detector fails to trigger, or the multi-bit payload is sufficiently corrupted to prevent metadata recovery. Crucially, this evasion must be achieved while preserving the visual quality of the original image, keeping perceptual distortion to a minimum. A secondary, opportunistic objective is \textit{ownership stealing} (or forgery). Since our attack operates by applying a second watermark, it naturally establishes a new provenance trace. 

\paragraph{Attacker Capabilities} We model a casual adversary representing an average internet user. Unlike sophisticated attackers considered in some prior work, they require no knowledge of watermarking internals, detection APIs, keys, or gradients. They only need access to off-the-shelf watermarking encoders---tools that are freely distributed and actively promoted, making this a highly realistic threat.

\paragraph{Attacker Knowledge} We evaluate this threat under two knowledge assumptions. In the \textit{beige-box} setting (\Cref{sec:policy}), the attacker knows which victim watermarking method was applied and selects the optimal attack watermark accordingly. In the \textit{black-box} setting (\Cref{sec:classify-wm,sec:pipeline}), the attacker has no prior knowledge of the victim method, relying instead on a trained classifier to identify the method from visual inspection alone before automatically deploying the correct attack.

\subsection{Experimental Setup} \label{sec:exp-setup}
To provide a consistent basis for the results presented in \Cref{sec:policy,sec:classify-wm,sec:pipeline}, we briefly outline the shared experimental methodology, detailing any section-specific variations later.

\paragraph{Datasets, Methods \& Metrics} We evaluate our attacks on subsets of high-quality images from the DiffusionDB \cite{wang2023diffusiondb} and MS-COCO \cite{lin2014microsoft} datasets (filtering details in \Cref{app:setup_details}). We consider eight representative watermarking schemes (see \Cref{tab:waterm_methods_overview} of the Appendix): two in-processing (Stable Signature, Tree-Ring) and six post-processing (StegaStamp, RoSteALS, ZoDiac, Pixel Seal, WAM, Video Seal). 
We evaluate attack success via two criteria: detection evasion (measured by True Positive Rate at a 1\% False Positive Rate, TPR@1\%FPR) and payload corruption (measured by Bit Accuracy, BA). Crucially, many multi-bit methods (like StegaStamp and RoSteALS) rely heavily on Error-Correcting Codes (ECC). Once bit accuracy drops below their correction capacity (typically 90-92\%), decoding fails catastrophically \cite{Tancik_2020_CVPR, bui2023rosteals}. We therefore track raw bit accuracy to capture these critical failure points. For 0-bit methods like ZoDiac and Tree-Ring, which do not carry a recoverable message, we additionally track the shift in their respective detection statistics (the $\ell_1$-distance and $p$-value, respectively) to provide a complete picture of signal degradation. Finally, perceptual distortion is evaluated using a composite metric, Normalized Quality Degradation (NQD), which aggregates eight standard image quality metrics (details in \Cref{app:calculate_nqd}).

\paragraph{Baselines} We compare our attack against the most effective generic removal attacks identified in the WAVES benchmark \cite{an2024waves}: prompted diffusion regeneration, VAE regeneration, and iterative diffusion ``rinsing'' (hyperparameters detailed in \Cref{app:setup_details}). Crucially, prompted diffusion regeneration assumes the attacker has access to the exact text prompt originally used to generate the image. This is a highly restrictive assumption that rarely holds for third-party internet users. By contrast, our attack requires only the image itself, making it a substantially more viable threat in the wild.

\section{Beige-Box Interference: Deriving the Attack Policy} \label{sec:policy}

To build an effective, automated removal attack, we must first understand how different watermarking schemes interact when superimposed. This section serves as the first stage in developing our pipeline. By systematically mapping watermark interference in a beige-box setting, we aim to derive a simple decision rule: given a known victim watermark, select the attack watermark that maximizes erasure while minimizing perceptual distortion. 

\paragraph{Policy Training Procedure} To map this vulnerability space and learn the optimal policy, we define a data-driven procedure. Given a dataset of training images $\mathcal{D}$, a set of victim watermarking schemes $\mathcal{V}$, and a set of attack watermarking schemes $\mathcal{A}$, the procedure is as follows: For each victim method $v \in \mathcal{V}$ and each image $x \in \mathcal{D}$, we apply the embedder of $v$. During embedding, we randomize all available parameters (e.g., messages, cryptographic keys, and seeds) to ensure robust sampling of the watermark distribution. Next, for each attack method $a \in \mathcal{A}$, we apply the embedder of $a$ directly on top of the victim-watermarked images, again utilizing random parameters. Finally, we run the corresponding victim detector $v$ on these twice-watermarked images. For every victim-attack pair $(v, a)$, we aggregate the results over the entire dataset $\mathcal{D}$ to compute specific performance metrics: detection evasion (measured via TPR@1\%FPR), payload corruption (measured via bit accuracy), and perceptual image quality degradation. Executing this across all combinations yields a $|\mathcal{V}| \times |\mathcal{A}|$ matrix of interference effects, from which the optimal attack action can be selected.

\begin{figure}[t]
    \centering
    \includegraphics[width=\linewidth, trim={0.35cm 0.7cm 0.7cm 0cm}, clip]{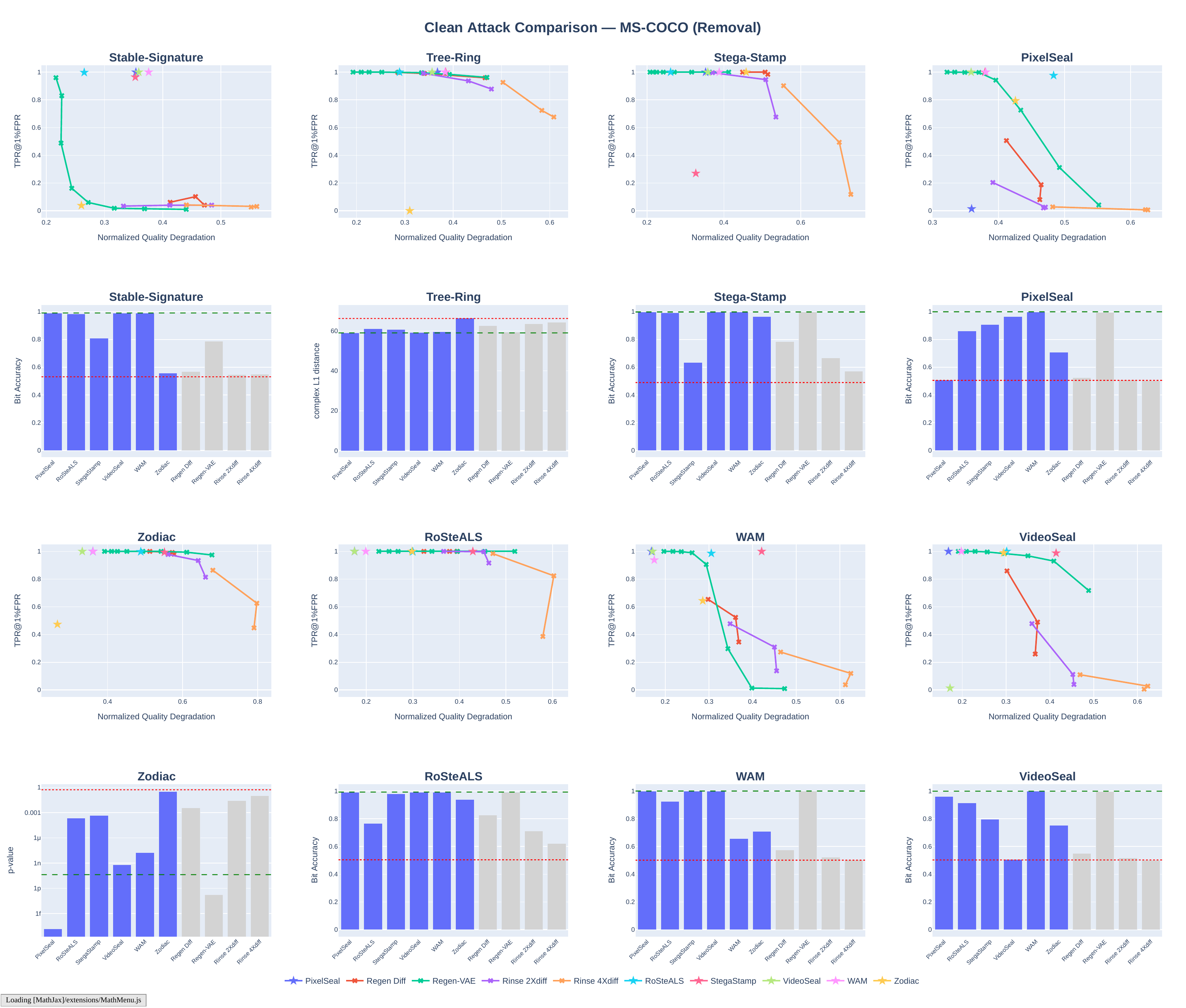}
    \caption{
    Attack evaluation on MS-COCO (see \Cref{fig:waw_big_figure_diffdb} of the Appendix for DiffusionDB results). Rows~1 and~3 plot detection rate (TPR@1\% FPR, lower is better) vs.\ image quality degradation (NQD, lower is better). In these plots, re-watermarking attacks are denoted by stars and WAVES baselines by crosses. 
    Rows~2 and~4 plot the bit accuracy (higher is better) for the message-based methods, $\ell_1$-distance for Tree-Ring (higher is better) and the $p$-value for ZoDiac (lower is better).
    In these plots, re-watermarking attacks are shown as blue bars and baselines as gray bars.
    }
    \label{fig:waw_big_figure}
\end{figure}

\paragraph{Training Setup} We instantiate this methodology using the watermarking methods, datasets and metrics defined in \Cref{sec:threat_model}. 
Since in-processing methods (Stable Signature, Tree-Ring) require access to the generative process, they cannot be applied post-hoc to existing images. Therefore, the available attack set $\mathcal{A}$ is restricted to the six post-processing methods.
\Cref{tab:attack-matrix} presents a visual summary of the resulting interference matrix learned during this phase, with detailed metric breakdowns shown in \Cref{fig:waw_big_figure}.

\begin{table}[b]
  \begin{minipage}{0.5\textwidth}
    \setlength{\tabcolsep}{2pt}
    \footnotesize
    
    \renewcommand{\arraystretch}{1.1}
    
    \newcommand{\rot}[1]{\multicolumn{1}{c}{\makebox[1.5em][l]{\raisebox{-1em}[0pt][0pt]{\rotatebox[origin=bl]{35}{#1}}}}}
    
    \begin{tabular}{l | *{8}{>{\centering\arraybackslash}m{1.7em}}}
     
     \multicolumn{1}{l|}{\diagbox[width=5.5em, height=3em]{\textbf{Attack}}{\textbf{Victim}}} & 
     \rot{Stable Sig.} & \rot{Tree-Ring} & \rot{StegaStamp} & \rot{Pixel Seal} & 
     \rot{ZoDiac} & \rot{RoSteALS} & \rot{WAM} & \rot{Video Seal} \\
    \hline

     StegaStamp   & \cellcolor{miniaccentblue} & \cellcolor{lightblue} & \cellcolor{accentlightpink}\crossmark & \cellcolor{lightpink} & \cellcolor{miniaccentpink} & \cellcolor{lightpink} & \cellcolor{lightpink} & \cellcolor{miniaccentpink} \\
     Pixel Seal & \cellcolor{lightblue} & \cellcolor{lightblue} & \cellcolor{lightpink} & \cellcolor{accentlightpink}\crossmark & \cellcolor{lightpink} & \cellcolor{lightpink} & \cellcolor{lightpink} & \cellcolor{lightpink} \\
     ZoDiac     & \cellcolor{accentlightblue}\crossmark & \cellcolor{accentlightblue}\crossmark & \cellcolor{lightpink} & \cellcolor{miniaccentpink} & \cellcolor{accentlightpink}\crossmark & \cellcolor{lightpink} & \cellcolor{miniaccentpink} & \cellcolor{lightpink} \\
     RoSteALS   & \cellcolor{lightblue} & \cellcolor{lightblue} & \cellcolor{lightpink} & \cellcolor{lightpink} & \cellcolor{miniaccentpink} & \cellcolor{miniaccentpink}\crossmark & \cellcolor{lightpink} & \cellcolor{lightpink} \\
     WAM        & \cellcolor{lightblue} & \cellcolor{lightblue} & \cellcolor{lightpink} & \cellcolor{lightpink} & \cellcolor{lightpink} & \cellcolor{lightpink} & \cellcolor{accentlightpink}\crossmark & \cellcolor{lightpink} \\
     Video Seal & \cellcolor{lightblue} & \cellcolor{lightblue} & \cellcolor{lightpink} & \cellcolor{lightpink} & \cellcolor{lightpink} & \cellcolor{lightpink} & \cellcolor{lightpink} & \cellcolor{accentlightpink}\crossmark \\
    \end{tabular}
  \end{minipage}
  \hfill
  \begin{minipage}{0.45\textwidth}
    \caption{Matrix of interference effects between victim (columns) and attack (rows) watermarks. Cell shading reflects attack strength: no effect (light), partial degradation (medium) and watermark effectively removed (dark). A cross (\crossmark) marks the attack selected by our derived policy (re-application for post-processing methods; ZoDiac for in-processing methods).}
    \label{tab:attack-matrix}
  \end{minipage}
\end{table}

\paragraph{In-processing Victims} We find that in-processing watermarks can be effectively erased by applying ZoDiac on top. ZoDiac is the only attack watermark to reduce TPR to near zero for both Tree-Ring and Stable Signature. Crucially, it achieves this while introducing far less perceptual degradation than any of the diffusion-based baseline attacks (compare the yellow ZoDiac star to the baseline crosses in \Cref{fig:waw_big_figure}, Rows 1 \& 3).
Other post-processing methods have limited impact, establishing ZoDiac as the optimal response for in-processing victims.

\paragraph{Post-processing Victims} For post-processing watermarks, we find that the reapplication of the \emph{same} method is consistently the most destructive attack. StegaStamp, Pixel Seal, and Video Seal are completely overwritten by a second embedding of themselves, reducing TPR to near zero and bit accuracy to the unwatermarked baseline. 
\textbf{Crucially, re-watermarking dominates the baseline attacks in the evasion-quality trade-off:} it causes over $2\times$ less visual degradation than the strongest WAVES baselines while achieving equal or better erasure. This favorable trade-off is clearly visible in \Cref{fig:waw_big_figure} (Rows 1 \& 3), where same-method re-watermarking lies to the bottom-left of the baselines. 

ZoDiac and RoSteALS are the most robust victims overall, with no single attack watermark drives their detection below 50\% TPR.  However, since even the strongest WAVES baseline (Rinse $4\times$Diff) barely reaches this bound for RoSteALS, re-watermarking remains highly competitive. WAM presents an interesting case: ZoDiac aggressively reduces WAM's TPR %
but with high inter-image Bit-Accuracy variance, %
succeeding on some images while failing on others. Conversely, WAM-on-WAM achieves a higher mean TPR %
but suppresses bit accuracy %
more consistently across the dataset, making payload recovery impossible.
Finally, ZoDiac also shows partial cross-method effectiveness %
against Pixel Seal.
Further analysis of the quality degradation driving these dynamics is provided in \Cref{app:quality_degradation}.

\paragraph{The Attack Policy} Synthesizing these findings yields a straightforward, empirically grounded decision rule for our automated pipeline: \textbf{apply ZoDiac against in-processing victims, and reapply the identified method against post-processing victims.}

\paragraph{Ownership Forgery}
A key advantage of using watermarks as an attack tool %
is the ability to fulfill the adversary's secondary objective: ownership forgery (\Cref{sec:threat_model}). Re-watermarking does not merely degrade the victim's signal; it simultaneously embeds a new one. We evaluate forgery success here in the beige-box setting to determine if this new provenance trace is viable.

\definecolor{lightgray}{gray}{0.85}
\newcolumntype{g}{>{\columncolor{lightgray}}c}

\begin{table}[h]
\caption{Attack watermark recovery after embedding on top of an existing watermark. For each attack method, we report metrics averaged over both datasets for two scenarios: (1)~\emph{cross-method} (applied on top of different victims, averaged across all), and (2)~\emph{same-method} (re-applied on top of itself). %
}
\label{tab:attack_wm_waw}
\centering
\small
\begin{tabular}{lcccccccccccc}
\toprule
 & \multicolumn{2}{c}{\textbf{RoSteALS}} & \multicolumn{2}{c}{\textbf{StegaStamp}} & \multicolumn{2}{c}{\textbf{WAM}} & \multicolumn{2}{c}{\textbf{Video Seal}} & \multicolumn{2}{c}{\textbf{Pixel Seal}} & \multicolumn{2}{c}{\textbf{ZoDiac}} \\
\cmidrule(lr){2-3} \cmidrule(lr){4-5} \cmidrule(lr){6-7} \cmidrule(lr){8-9} \cmidrule(lr){10-11} \cmidrule(lr){12-13}
\textbf{Scenario} & BA & TPR & BA & TPR & BA & TPR & BA & TPR & BA & TPR & p-value &    TPR \\
\midrule
Cross  & 0.99 & 1 & 0.99 & 1 & 0.99 & 1 & 0.99 & 1 & 0.99 & 1 & 1.27e-8 & 1 \\
Same   & 0.83 & 1 & 0.99 & 1 & 0.78 & 0.69 & 0.99 & 1 & 0.99 & 1 & 1.76e-8 & 1 \\
\bottomrule 
\end{tabular}
\end{table}

As shown in \Cref{tab:attack_wm_waw}, in the cross-method setting, recovery is perfect across all attack methods, confirming the attacker's watermark is fully preserved. In the same-method setting, most methods still achieve perfect recovery, with only minor drops in bit accuracy for RoSteALS and WAM where structurally identical signals interfere. Crucially, the attack watermark's bit accuracy heavily dominates the victim's in all scenarios (compare to \Cref{fig:waw_big_figure}), meaning the attacker's message remains the dominant signal. This allows the attacker to reliably claim ownership or introduce sufficient ambiguity to contest the victim's provenance.

\section{Breaking Undetectability} \label{sec:classify-wm}

To effectively apply the attack policy established in \Cref{sec:policy} in a realistic black-box scenario, the attacker must first determine which watermarking method was used on the victim image.

A core design goal of modern image watermarking schemes is undetectability: watermarked images should be perceptually and statistically indistinguishable from their unwatermarked counterparts. If an adversary cannot even tell whether an image is watermarked, let alone determine which scheme was used, then targeted attacks become significantly harder. We challenge this assumption by demonstrating that a lightweight classifier suffices to %
identify the specific watermarking scheme from the image alone, without access to the detection model or embedding keys.

\paragraph{Classifier Architecture and Training}
We fine-tune a ConvNeXt-V2 Large model \cite{woo2023convnextv2}, pre-trained on ImageNet-22K \cite{deng2009imagenet}, as a 9-class classifier to distinguish %
the eight watermarking methods 
and unwatermarked images. We utilize 500 images per class from DiffusionDB \cite{wang2023diffusiondb} with a stratified 80/10/10 split, processing all inputs at a $512 \times 512$ resolution to preserve watermark artifacts. 
Full training details and hyperparameter configurations are provided in \Cref{app:classifier_ablations}.

\begin{wraptable}[21]{r}{0.38\textwidth}
\vspace{-2em}
\caption{Per-class recall of our primary classifier (cl1) on the in-distribution
DiffusionDB held-out test split and on out-of-distribution
MS-COCO images re-watermarked with each method
(500 images per class).}
\label{tab:clf_perclass}
\begin{center}
\small
\begin{tabular}{lcc}
\toprule
\textbf{Class} & \textbf{DiffDB} & \textbf{MS-COCO} \\
\midrule
Pixel Seal       & 0.960 & 0.930 \\
RoSteALS        & 1.000 & 0.998 \\
Stable Sig.     & 1.000 & 0.992 \\
StegaStamp      & 1.000 & 1.000 \\
Tree-Ring       & 0.780 & 0.728 \\
WAM             & 1.000 & 0.992 \\
ZoDiac          & 0.980 & 0.634 \\
Video Seal       & 1.000 & 0.968 \\
Unwaterm.   & 0.860 & 0.660 \\
\midrule
Overall acc.    & 0.953 & 0.878 \\
Macro-F1        & 0.953 & 0.876 \\
\bottomrule
\end{tabular}
\end{center}
\end{wraptable}

\paragraph{Classifier Results} We evaluate the classifier on a held-out DiffusionDB split (in-distribution) and an unseen set of MS-COCO images watermarked with each method (out-of-distribution). As shown in \Cref{tab:clf_perclass}, the classifier achieves 95.3\% accuracy on the in-distribution split, identifying six of the eight methods with near-perfect recall.

Crucially, the classifier generalizes well to the unseen MS-COCO dataset: overall accuracy drops by less than eight percentage points, %
with most methods retaining recall above 96\%.
While frequency-based methods like Tree-Ring and ZoDiac are slightly harder to identify under distribution shift, the classifier still routes the majority of these images correctly.
Furthermore, classification performance remains largely preserved on images from a different diffusion backbone (SD~3.5; \Cref{app:classifier_sd35}). %
\Cref{app:classifier_distribution} provides further analysis, including common inter-class confusions.

\paragraph{Handling Misclassifications}  The weakest class under distribution shift is ``unwatermarked,'' where the model misclassifies roughly a third of clean images as watermarked. In our threat model, this is an acceptable failure mode. The classifier's role is not to serve as a high-precision production detector but to route images to the best attack policy. A false positive on an unwatermarked image merely triggers an unnecessary re-watermarking step; this slightly degrades the image's quality but causes no security harm to the attacker. Recall on watermarked classes would likely improve with a more diverse training set, which an attacker could assemble cheaply. 

\paragraph{Implications for Undetectability}
Our setup is deliberately minimal: we use an off-the-shelf backbone, a single small dataset, no watermark-specific feature engineering, no model ensembling. 
The ease with which this classifier succeeds suggests that the artifacts left by current watermarking schemes are far more distinctive than their undetectability claims imply.
Alternative training recipes are evaluated in \Cref{app:classifier_ablations}, but do not alter this overall conclusion. Ultimately, this lightweight classifier is more than sufficient to automate the attack policy derived in \Cref{sec:policy}.

\section{The End-to-End Attack Pipeline} \label{sec:pipeline}

\begin{table}[t]
\caption{Decoder metrics on MS-COCO for our attack vs.\ baselines. BA\textsuperscript{clean}, BA\textsuperscript{unwm}, BA\textsuperscript{atk} denote the bit accuracy on clean watermarked images, unwatermarked images and attacked watermarked images, respectively. TPR is reported at 1\%FPR and \#mis denotes the number of images misclassified as unwatermarked. Bold indicates the strongest attack metric achieved per victim method. Effective attacks yield lower BA and TPR, except for Tree-Ring (\textsuperscript{*}$\ell_1$ distance) and ZoDiac (\textsuperscript{**}$p$-value) where higher is better. Baseline strengths are chosen to not exceed our attack's quality degradation, though Rinse 4$\times$Diff still exceeds it even at the lowest strength. See \Cref{tab:pipeline_tab_diffdb} in the Appendix for DiffusionDB.}
\label{tab:pipeline_tab}
\centering
\footnotesize
\setlength{\tabcolsep}{4pt}
\resizebox{\textwidth}{!}{%
\begin{tabular}{lcccccp{0.05cm}ccp{0.05cm}ccp{0.05cm}ccp{0.05cm}cc}
\toprule
&&& \multicolumn{3}{c}{\textbf{Our Attack}} & & \multicolumn{2}{c}{\textbf{Regen-Diff 100}} & & \multicolumn{2}{c}{\textbf{RegenVAE 2}} & & \multicolumn{2}{c}{\textbf{Rinse 2$\times$Diff 20}} & & \multicolumn{2}{c}{\textbf{Rinse 4$\times$Diff 20}} \\
\cmidrule(lr){4-6} \cmidrule(lr){8-9} \cmidrule(lr){11-12} \cmidrule(lr){14-15} \cmidrule(lr){17-18}
\textbf{Method} & BA\textsuperscript{clean} & BA\textsuperscript{unwm} & BA\textsuperscript{atk} & TPR & \#mis & & BA\textsuperscript{atk} & TPR & & BA\textsuperscript{atk} & TPR & & BA\textsuperscript{atk} & TPR & & BA\textsuperscript{atk} & TPR \\
\midrule
Pixel Seal  & 0.999 & 0.505 & 0.514 & {0.029} & 9  & &  {0.524} &  {0.08} & &  {0.554} &  {0.312} & &  {0.542} &  {0.204} & & \textbf{0.512} & \textbf{0.028} \\

RoSteALS   & 0.994 & 0.504 & \textbf{0.749} & {0.992} & 0  & &  {0.826} &  {1.000} & &  {0.9587} &  {1.000} & &  {0.868} &  {1.000} & &  {0.761} & \textbf{0.984} \\

Stable Sig.  & 0.981 & 0.531 & \textbf{0.535} & {0.039} & 0  & & {0.568} & {0.041} & & {0.569} & \textbf{0.014} & & {0.562} & {0.034} & & {0.562} & {0.042} \\

StegaSt    & 0.998 & 0.509 & \textbf{0.592} & \textbf{0.222} & 0  & & {0.784} & {0.984}& &{0.994} &{1.000} & &{0.870} &{0.996} & &{0.749} & {0.902}\\

Tree-Ring   & 53.3\textsuperscript{*} & 66.2\textsuperscript{*} & \textbf{64.8\textsuperscript{*}} & \textbf{0.280} & 62 & & {62.6\textsuperscript{*}} & {0.96} && {62.1\textsuperscript{*}}  & {0.984}& & {62.4\textsuperscript{*}} & {0.992} && {63.2\textsuperscript{*}} & {0.926} \\

Video Seal  & 0.998 & 0.503 & \textbf{0.520} & \textbf{0.046} & 3  & & {0.550} & {0.260} & & {0.697}&{0.930} & & {0.575}&{0.479} & &{0.533} & {0.110}\\

WAM        & 0.999 & 0.501 & {0.707} & {0.862} & 0  & & {0.576}& {0.346}& &\textbf{0.501} & \textbf{0.014} & &{0.612} & {0.478}& &{0.552} & {0.274}\\
ZoDiac     & 3.8e-11\textsuperscript{**} & 5.3e-1\textsuperscript{**} & \textbf{3.3e-1\textsuperscript{**}} & \textbf{0.464} & 18 & & {3.5e-3\textsuperscript{**}} & {0.980} & & {1.8e-3\textsuperscript{**}}& {0.994} & & {3.3e-3\textsuperscript{**}}& {0.978}& & {1.7e-2\textsuperscript{**}}& {0.864}\\
\bottomrule
\end{tabular}}
\end{table}

\begin{wrapfigure}[18]{r}{0.5\textwidth}
\vspace{-0.6cm}
    \centering
    \includegraphics[width=\linewidth]{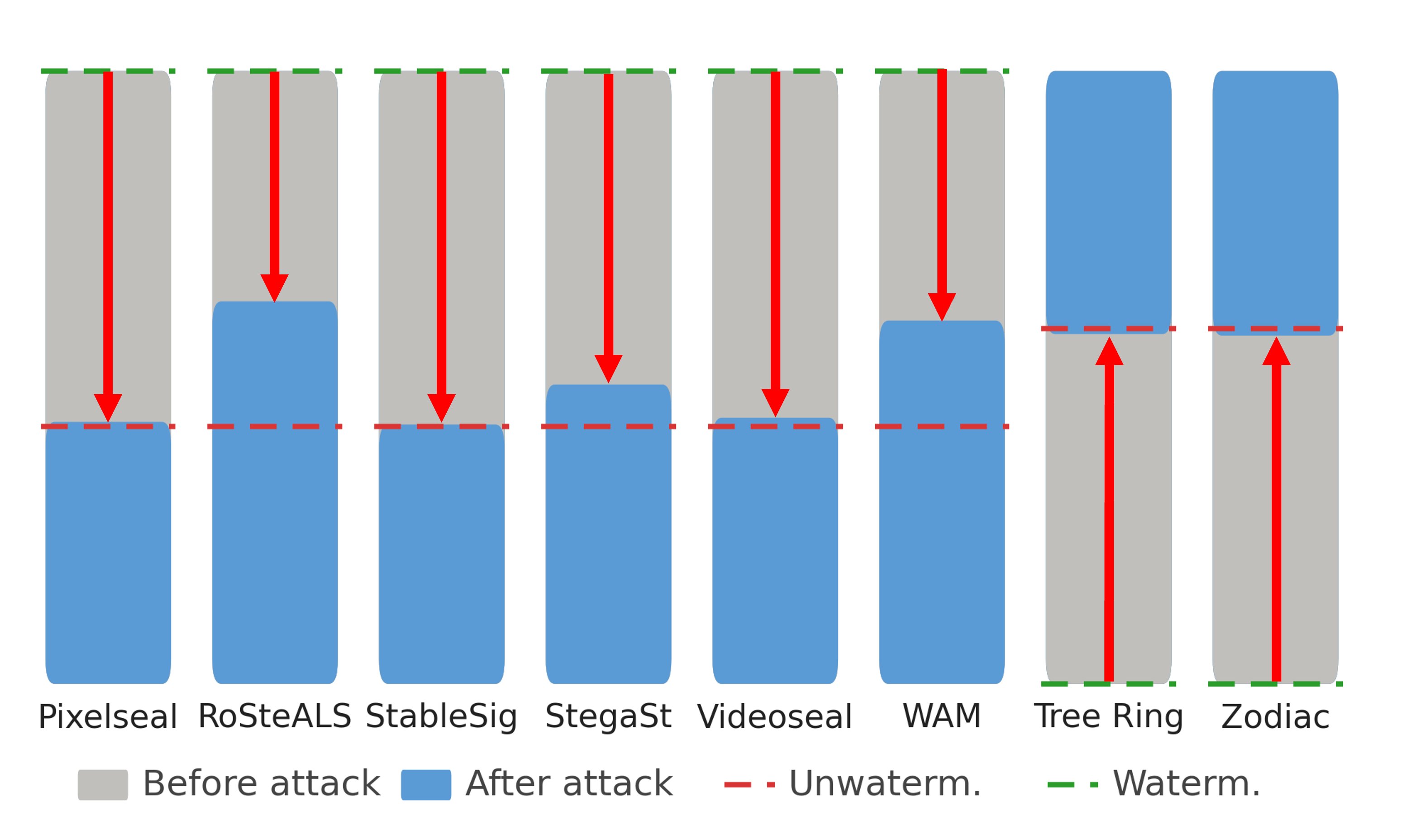}
    \caption{%
        Detection metrics before and after applying the end-to-end WAW attack pipeline. We report mean bit accuracy for multi-bit methods (where higher indicates successful detection). For 0-bit methods, we report the relevant detection statistic: $\ell_1$-distance for Tree-Ring and $p$-value for ZoDiac (where lower indicates successful detection).
    }
    \label{fig:pipeline_attack_vis}
\end{wrapfigure}

We now introduce the Watermarks Attack Watermarks (WAW) pipeline: an automated, end-to-end attack that synthesizes the capabilities developed in the preceding sections. We evaluate this pipeline in a realistic \textit{black-box} setting, where the attacker has no prior knowledge of which watermarking method was used to protect the victim image.

\paragraph{Pipeline Execution} The WAW pipeline operates in two steps. First, for each watermarked image, it applies the classifier from \Cref{sec:classify-wm} to predict the underlying watermarking algorithm from visual inspection alone. Second, based on this prediction, it executes the deterministic attack policy derived in \Cref{sec:policy}: if the image is classified as containing an in-processing watermark (Tree-Ring, Stable Signature) or ZoDiac, the pipeline applies ZoDiac on top. For all other post-processing methods, it reapplies the predicted method itself. We evaluate this automated pipeline on both the MS-COCO and DiffusionDB datasets, reporting the detailed metrics in Tables \ref{tab:pipeline_tab} and \ref{tab:pipeline_tab_diffdb} and presenting a visual summary of the signal degradation in \Cref{fig:pipeline_attack_vis}. \textbf{Crucially, the WAW pipeline outperforms or effectively matches the baseline attacks across all methods except WAM}---either by achieving strictly lower detection rates or by more aggressively corrupting the embedded payload (as with RoSteALS)---while operating at a comparable or lower perceptual cost.

\paragraph{Evaluation Results} Since our classifier achieves high accuracy, the WAW pipeline remains highly successful, causing a severe drop in the victim's embedded signal across most methods. For half of the evaluated schemes, bit accuracy drops entirely to the unwatermarked baseline with TPR close to zero. For highly robust methods like RoSteALS and WAM, TPR remains higher, but bit accuracy still falls below 75\%, meaning neither can successfully recover their embedded payload. Overall, the black-box pipeline's performance closely mirrors the ``oracle'' beige-box attack performance from \Cref{sec:policy}.

\paragraph{Impact of Misclassifications} The methods that show a slight performance drop compared to the beige-box oracle are, as expected, precisely those where the classifier is weakest. For Tree-Ring, the TPR rises to 28\% because a significant number of images (62 on MS-COCO) are incorrectly classified as ``unwatermarked.'' Consequently, the pipeline bypasses the attack, leaving the original watermark intact.

A similar but milder effect occurs for ZoDiac. Notably, a high number of ZoDiac images (142~on MS-COCO) are instead misclassified as Tree Ring, and interestingly, a high number of Tree-Ring images (72~on MS-COCO) are misclassified as ZoDiac. However, because our policy prescribes ZoDiac as the optimal attack for both, the pipeline preserves high removal performance. Finally, the pipeline achieves slightly better attack performance on DiffusionDB than MS-COCO. We attribute this to DiffusionDB matching the classifier's training distribution, though the evaluation images were strictly held-out.

\paragraph{Beyond Re-Watermarking}
We emphasize that the WAW attack strategy is only one way to exploit method identification. Our pipeline uses the classifier purely to select the most effective overwriting attack. However, an adversary empowered with the knowledge of the victim scheme could additionally mount highly sophisticated, method-specific attacks (e.g., targeted perturbation of specific frequency bands or surrogate-detector optimization) that exploit the individual vulnerabilities of each scheme, possibly achieving even stronger removal \cite{shamshad2025firstplacesolutionneurips2024, lin2025crack}. Beyond direct removal, the classifier can also serve as a low-cost oracle to verify whether an attack attempt (both removal or forgery) was successful, without requiring access to the victim's proprietary detector. We expand on these vulnerabilities in the Broader Impacts section.

\section{Discussion \& Conclusion}

Watermark encoders are, almost by construction, effective tools for attacking other watermarks. Re-watermarking (simply applying a second watermark on top of an existing one) reliably suppresses the victim's embedded signal across all eight watermarking methods we consider, including both post-processing and in-processing schemes, while introducing far less perceptual degradation than diffusion-based removal attacks. Crucially, re-watermarking does not only erase: it simultaneously embeds a new, recoverable message, allowing the attacker to plausibly claim ownership while undermining the original provenance. From our experiments, we derive a simple attack policy: for in-processing watermarks, re-watermark with ZoDiac; for post-processing watermarks, reapply the method itself. Despite its simplicity, this policy achieves strong erasure across all victim methods.

A natural objection to this attack is that it requires knowing which watermarking method was used. We have shown this assumption is easy to drop: a lightweight classifier trained on a small dataset identifies the watermarking method from the image alone with high accuracy, closing the loop into a fully automated, blind attack pipeline. The success of this classifier confirms that the different methods we study leave scheme-specific fingerprints that are recoverable without any key or model access, which opens the door to the usage of method-specific attacks. 

Mitigation is challenging due to an inherent tension in watermark design. To survive common distortions, a watermark must introduce a sufficiently strong signal. However, if this signal were truly undetectable, it would be indistinguishable from natural image noise. We demonstrate that modern schemes fail these criteria in two ways: first, their signals are distinct enough to be learned by a simple classifier; second, their robustness mechanisms are highly brittle to the structured perturbations introduced by other watermarking encoders. As long as watermarks rely on specific, non-overlapping frequency or spatial artifacts to achieve robustness, they will remain vulnerable to being identified and overwritten by competing schemes operating under the same perceptual constraints.

Taken together, our results suggest that the robustness and imperceptibility properties that make image watermarks attractive are also what makes them effective attack tools against each other. We hope this reframing motivates the development of watermarking schemes with stronger formal security guarantees and more emphasis on undetectability, and that our pipeline serves as a cheap, reproducible stress-test for evaluating candidate schemes before deployment.

\section*{Acknowledgements}
This research was supported by The University of Melbourne’s Research Computing Services and the Petascale Campus Initiative. We also acknowledge the assistance of Claude, Perplexity, and GitHub Copilot for scaffolding code in the released supplementary material.

{
    \small
    \bibliographystyle{abbrv}
    \bibliography{main}
}

\appendix
\crefalias{section}{appendix}      %
\crefalias{subsection}{appendix}   %
\crefalias{subsubsection}{appendix} %
\section*{Broader Impacts} \label{sec:broader_impacts}

This work exposes a previously underappreciated attack surface in invisible image 
watermarking: that watermark encoders themselves can be repurposed as cheap, effective 
removal tools, and that the identity of the watermarking method can be inferred from 
visual inspection alone. We discuss both the positive and negative societal implications 
of these findings.

\paragraph{Positive Impacts} Invisible watermarking is increasingly relied upon as a 
technical mechanism for provenance assertion and deepfake detection, including under 
emerging regulatory frameworks such as the EU AI Act. Our results demonstrate that 
several widely deployed schemes are far more fragile than their stated design goals 
suggest, and that the assumption of method secrecy (a common implicit defense) offers 
little practical protection. This has direct value for regulators and platforms that 
must select watermarking schemes for deployment: our pipeline provides a cheap and 
reproducible stress-test that can expose fragile schemes before they are mandated or 
relied upon in production systems. More broadly, our classifier corroborates recent 
theoretical results showing that all prior image watermarking schemes leave detectable 
traces~\cite{gunn2025undetectable}, strengthening the case for adopting 
cryptographically grounded undetectability as a first-class design requirement rather 
than an implicit assumption. Finally, the same classifier that an adversary could use 
to verify removal success can equally be deployed by watermark providers to audit 
whether their scheme leaves scheme-specific fingerprints recoverable without key 
access, turning a potential attack tool into a diagnostic instrument for defenders. 
By making these vulnerabilities explicit and reproducible, we hope to motivate the 
development of watermarking schemes with stronger formal security guarantees. 

\paragraph{Negative Impacts} We acknowledge that the attack pipeline described in this 
paper could be misused. First, an adversary could use our method to 
strip copyright or provenance watermarks from AI-generated content, undermining 
ownership claims and complicating the enforcement of content authenticity policies. 
Second, the classifier enables method identification, which in turn facilitates 
method-specific beige-box attacks that go beyond our re-watermarking pipeline: as 
discussed in \Cref{sec:intro}, knowing the watermarking scheme allows an adversary to 
exploit its precise technical vulnerabilities, achieving substantially stronger removal 
than generic black-box strategies. Third, the classifier can serve as a low-cost 
oracle for verifying attack or forgery success: by checking whether a processed image is 
classified as unwatermarked, an adversary can confirm whether their attack attempt 
was effective without access to the victim's detector. The barrier to entry for all of the above is low, as the attack requires only publicly available watermarking encoders and no 
access to detection keys, surrogate models, or gradients. We note, however, that all 
components of our pipeline are already publicly available, and that the fundamental 
vulnerability exists independently of this work. Our contribution is to make the 
vulnerability legible and to demonstrate its practical severity. We therefore believe that the value of disclosure outweighs the marginal increase in misuse risk.

\section{Extended Related Work}\label{app:related}

\paragraph{Collision and Ambiguity Attacks}
Collision and ambiguity have been proposed as attack mechanisms in related settings. ``Watermark collision'' for logit‑based LLM watermarks~\cite{luo2025lost} introduced the idea that embedding multiple watermarks in the same text can significantly degrade detection accuracy. The authors propose collision as a general attack philosophy and speculatively suggest that an existing watermark detector could be repurposed to drive such an attack---but they do not pursue this direction further or test it empirically. Our work can be seen as realizing this idea in the image domain: we confirm that watermarks do interfere with one another, and we show that a learned classifier can identify which watermark is present and select the most damaging one to apply. Ambiguity attacks on model watermarks~\cite{chen2023effective} consider an adversary forging a substitute watermark and embedding it on top of the original, so that ownership claims become ambiguous since both watermarks may remain detectable. Crucially, their goal is not erasure but forging a competing ownership claim---a fundamentally different objective from ours.
\paragraph{Formal Undetectability}
Undetectability has been formalized as a cryptographic property for watermarks: a watermark is undetectable if no efficient adversary without the detection key can distinguish watermarked from unwatermarked outputs~\cite{christ2024undetectable}. \cite{gunn2025undetectable} instantiate this guarantee for generative image models and show empirically that all tested schemes leave detectable traces. Our classifier corroborates this finding: trained purely on visual features, it reliably identifies which watermarking method was applied, confirming that the schemes we study leave scheme-specific fingerprints recoverable without any key or model access. Formal undetectability thus remains an aspiration rather than a property of widely deployed schemes.
\paragraph{Attack Robustness}
Research into the robustness of digital watermarks has identified a variety of removal strategies, ranging from classical distortions (rotation, resizing, JPEG compression) to sophisticated adversarial and generative approaches. The WAVES benchmark~\cite{an2024waves} established a comprehensive stress-testing protocol involving 26 distinct attack variants, finding that regeneration-based attacks~\cite{zhao2024invisible, saberi2024robustness} are among the most consistently effective. Regen-Diff transfers a watermarked image into a latent representation and restores it using a surrogate diffusion model such as SD~1.4~\cite{zhao2024invisible}; Rinse-Diff subjects the image to multiple iterative cycles of noising and denoising~\cite{an2024waves, zhao2024invisible}; and Regen-VAE passes the image through a pre-trained variational autoencoder, utilizing the bottleneck to perturb the latent feature space~\cite{saberi2024robustness, balle2018variational}. The effectiveness of these generic strategies is often surpassed by method-specific beige-box attacks, which leverage precise knowledge of the watermarking algorithm to exploit its unique technical flaws~\cite{shamshad2025firstplacesolutionneurips2024, an2024waves}. Results from the NeurIPS 2024 ``Erasing the Invisible'' challenge confirm that beige-box attacks achieve significantly stronger removal than black-box ones~\cite{shamshad2025firstplacesolutionneurips2024, serzhenko2025watermark}: knowledge of the embedding mechanism enables targeted interventions such as surrogate-detector attacks or phase perturbations that cause substantially larger drops in detection accuracy~\cite{lin2025crack}.

\section{Watermarking Method Details}
\label{app:method_details}

In our experiments, we evaluate eight distinct watermarking algorithms, covering both major watermarking types: in-processing and post-processing. 

\paragraph{In-processing Watermarks} These methods adapt generative models to embed signatures during image creation, allowing the watermark to be deeply embedded within the image’s semantic structure. For example, Tree-Ring Watermarking \cite{tree_rings} embeds a Fourier pattern into the initial noise vector used for diffusion sampling. Alternatively, Stable Signature \cite{Fernandez_2023_ICCV} utilizes partial model modification by fine-tuning only the latent decoder of a pretrained diffusion model. Because these methods require access to the generative process, they cannot be applied to arbitrary existing images and thus serve only as victim watermarks in our evaluation.

\paragraph{Post-processing Watermarks} These methods are applied to images after they have been generated. They are model-agnostic and can be applied to any digital image. While StegaStamp \cite{Tancik_2020_CVPR} utilizes a deep CNN-based encoder-decoder architecture trained with differentiable perturbations for physical-world robustness, ZoDiac \cite{zhang2024attackresilient} and RoSteALS \cite{bui2023rosteals} focus on latent space integrity. ZoDiac optimizes the trainable latent space of a pretrained diffusion model for existing images, while RoSteALS employs a lightweight encoder to map secrets into the latent space of a frozen autoencoder. Advanced methods like Watermark Anything (WAM) \cite{sander2025watermark} introduce transformer-based architectures to enable localized detection and multiple message extraction. Finally, Pixel Seal and Video Seal \cite{soucek2025pixel} utilize adversarial-only training and Just Noticeable Difference attenuation to embed signatures into images while ensuring temporal consistency. An overview of all evaluated methods is provided in Table \ref{tab:waterm_methods_overview}.

\begin{table}[ht]
\caption{Watermarking methods overview. References under method names point to similar watermarking schemes represented by each method.}
\label{tab:waterm_methods_overview}
\centering
\footnotesize
\resizebox{\textwidth}{!}{%
\begin{tabular}{m{0.02\linewidth}|l|p{0.35\linewidth}|p{0.3\linewidth}|p{0.15\linewidth}|}
\cline{2-5}
 & \textbf{Method} & \textbf{What is Embedded} & \textbf{Detection Procedure} & \textbf{Metric} \\
\cline{2-5}
\multirow{4}{*}{\colorbox{lightpink}{\rotatebox[origin=c]{90}{\hspace{0.15cm}In-processing\hspace{0.15cm}}}}
& \parbox[t]{1.3cm}{Tree-Ring\\ \scriptsize{sim. \cite{ci2025wmadapter}}}
& Concentric ring pattern in Fourier frequency domain of latent noise
& DDIM inversion $\rightarrow$ FFT of recovered latent $\rightarrow$ L1 distance to expected ring pattern in frequency mask
& L1 distance (lower indicates watermarked) \\
\cline{2-5}
& \parbox[t]{1.3cm}{Stable\\Signature\\\scriptsize{sim. \cite{rebuffi2026learning}}}
& 48-bit binary message injected via finetuned Stable Diffusion image decoder
& HiddenDecoder extracts bits from pixel space
& \multirow{12}{*}{\parbox{\linewidth}{\centering Bit Accuracy (fraction of correctly decoded bits; higher indicates watermarked)}} \\
\cline{2-4}
\multirow{12}{*}[-0.08cm]{\colorbox{lightblue}{\rotatebox[origin=c]{90}{\hspace{1.6cm}Post-Processing\hspace{1.6cm}}}}
& \parbox[t]{1.3cm}{StegaStamp\\ \scriptsize{sim. \cite{yakushev2022docmarking, liang2025screenmark}}}
& 7-character ASCII text, BCH-encoded to 100 bits, embedded in pixel space
& StegaStamp decoder $\rightarrow$ BCH decoding $\rightarrow$ message reconstruction
& \\
\cline{2-4}
& \parbox[t]{1.3cm}{RoSteALS\\ \scriptsize{sim. \cite{meng2025latent, rezaei2024lawa}}}
& 7-character ASCII text, ECC-encoded to 100 bits, embedded via LDM latent manipulation
& RoSteALS decoder network $\rightarrow$ ECC decoding $\rightarrow$ message reconstruction
& \\
\cline{2-4}
& \parbox[t]{1.3cm}{WAM\\ \scriptsize{sim. \cite{yuan2024semi}}}
& 32-bit binary message, spatially distributed (network predicts per-pixel bits + watermark mask)
& WAM detector: spatial mask prediction + per-pixel bit decoding
& \\
\cline{2-4}
& \parbox[t]{1.3cm}{Pixel Seal\\ \scriptsize{sim. \cite{zhu2018hidden, bui2025trustmark}}}
& 96-bit binary message embedded in pixel space
& \multirow{2}{\linewidth}{Pixel/Video Seal detector $\rightarrow$ binary bit logits}
& \\
\cline{2-3}
& \parbox[t]{1.3cm}{Video Seal\\ \scriptsize{sim. \cite{zhang2019robust, chang2024dnn}}}
& 256-bit binary message embedded in pixel space
&
& \\
\cline{2-5}
& \parbox[t]{1.3cm}{ZoDiac\\ \scriptsize{sim. \cite{lei2025secure}}}
& Ring-shaped frequency pattern in VAE latent space, embedded via iterative DDIM optimization
& DDIM inversion $\rightarrow$ non-central chi-squared test in frequency domain
& p-value (lower indicates watermarked) \\
\cline{2-5}
\end{tabular}}
\end{table}

\section{Experimental Setup Details}
\label{app:setup_details}

\paragraph{Dataset Filtering} We conduct evaluations with images from the DiffusionDB \cite{wang2023diffusiondb} and MS-COCO \cite{lin2014microsoft} datasets. \cite{an2024waves} show that each of these datasets is characterized by a unique distribution of prompt words, where DiffusionDB emphasizes quality descriptors (e.g., ``beautiful,'' ``highly detailed''), whereas MS-COCO focuses on object descriptions. Following \cite{an2024waves}, we selected a subset of images as follows:
\begin{enumerate}
    \item Keep only the reference images and their prompt strings, discarding all other metadata.
    \item Tokenize prompts using OpenCLIP's tokenizer and keep only samples whose token count falls in $(0,75]$ (since longer prompts are truncated by Stable Diffusion \cite{rombach2022high}).
    \item Rank images by aesthetic score \cite{xu2023imagereward} and select the top 500 (prioritizing high-quality images for which watermarking is most practically relevant).
\end{enumerate}

\paragraph{Baseline Attack Configurations} As a baseline for the performance of our WAW attack, we evaluate each victim watermark against the strongest attacks identified in \cite{an2024waves}, each applied across a range of strengths:
\begin{itemize}
    \item \textbf{Prompted diffusion regeneration:} passes the image through Stable Diffusion~v1.4 conditioned on its original generative prompt (40--200 denoising timesteps).
    \item \textbf{VAE regeneration:} encodes and decodes the image via a pre-trained variational autoencoder \cite{balle2018variational} (quality levels 1--7).
    \item \textbf{Rinsing:} Regen-2xDiff and Regen-4xDiff iteratively noise and denoise the image via Stable Diffusion~v1.4 two and four times, respectively (20, 60, 100, and 10--50 timesteps per pass).
\end{itemize}

\section{Classifier Analysis and Ablation} \label{app:cl_big_section}

\subsection{Using a Different Diffusion Backbone (SD~3.5)}
\label{app:classifier_sd35}

We re-evaluate the classifier on images generated with the more recent model Stable Diffusion~3.5 \cite{esser2024scaling}. We regenerate 500 MS-COCO images with SD~3.5 and watermark them with each of the post-processing methods: Pixel Seal, RoSteALS,
StegaStamp, WAM, Video Seal and ZoDiac. For this experiment, Stable Signature and Tree-Ring are excluded by construction since both are trained for the specific SD~2.1-base backbone and would require substantial retraining to be swapped onto a different backbone. The classifier is the same model used in the
main paper without any additional fine-tuning or domain adaptation. Results can be found in \Cref{tab:clf_ablations} under MS-COCO (SD~3.5) where the column ``cl1'' represents the classifier presented in the main body of the paper. 

The post-processing watermarking methods remain largely identifiable under the backbone shift. RoSteALS, StegaStamp, WAM, and Video Seal all retain recall above 0.96, while Pixel Seal degrades only moderately (0.930~$\rightarrow$~0.824). The main degradation occurs in ZoDiac (recall: 0.634 $\to$ 0.490) and the unwatermarked class (0.660 $\to$ 0.070). ZoDiac embeds its watermark by DDIM-inverting the input image into the SD~2.1 latent space, overwriting a fixed spectral pattern in the latent representation, and reversing through SD~2.1 denoising while optimizing perceptual fidelity to the original. Unlike other post-processing methods, this process imprints SD~2.1-specific manifold statistics onto the watermarked image. Consequently, SD~3.5-generated images watermarked via ZoDiac exhibit a hybrid distribution—shifted toward SD~2.1 artifacts which seems to confound the classifiers decision boundary. The unwatermarked class failure similarly reflects the classifier's reliance on SD~2.1/DiffusionDB clean-image statistics, causing misclassification of SD~3.5 clean images (predominantly as Video Seal). However, as mentioned previously, such errors incur minimal quality cost in our attack-routing scenario due to unnecessary re-watermarking. The macro-F1 drop is thus attributable to these backbone-sensitive classes rather than watermark-method identification.

Figure~\ref{fig:cl3_distribution} shows the per-method prediction distribution of the cl1 classifier aggregated over MS-COCO and DiffusionDB. The classifier achieves near-perfect accuracy for Pixel Seal, RoSteALS, Stable Signature, StegaStamp, WAM, and Video Seal, confirming that these methods leave clear scheme-specific fingerprints that are easily recoverable. Notably, despite Pixel Seal and Video Seal sharing a learned encoder-decoder architecture, the classifier reliably distinguishes between them. The only notable confusion occurs between Tree-Ring and ZoDiac: approximately 70\% of Tree-Ring images are correctly classified, with the remainder misclassified predominantly as ZoDiac, and conversely around 20\% of ZoDiac images are misclassified as Tree-Ring. We attribute this mutual confusion to their shared embedding mechanism: both methods inject watermarks into the frequency spectrum of the diffusion latent during sampling and therefore seem to leave similar artifacts. As discussed in Section~\ref{sec:pipeline}, this confusion is operationally benign since both methods are routed to the same removal attack anyway. The only consequential failure mode is misclassification as \emph{Unwatermarked} (gray segment), visible for Tree-Ring, as such images bypass the attack pipeline entirely and their watermark is left intact. 

\subsection{Per-Method Classification Breakdown}
\label{app:classifier_distribution}

\begin{figure}[h]
    \centering
    \includegraphics[width=0.6\linewidth]{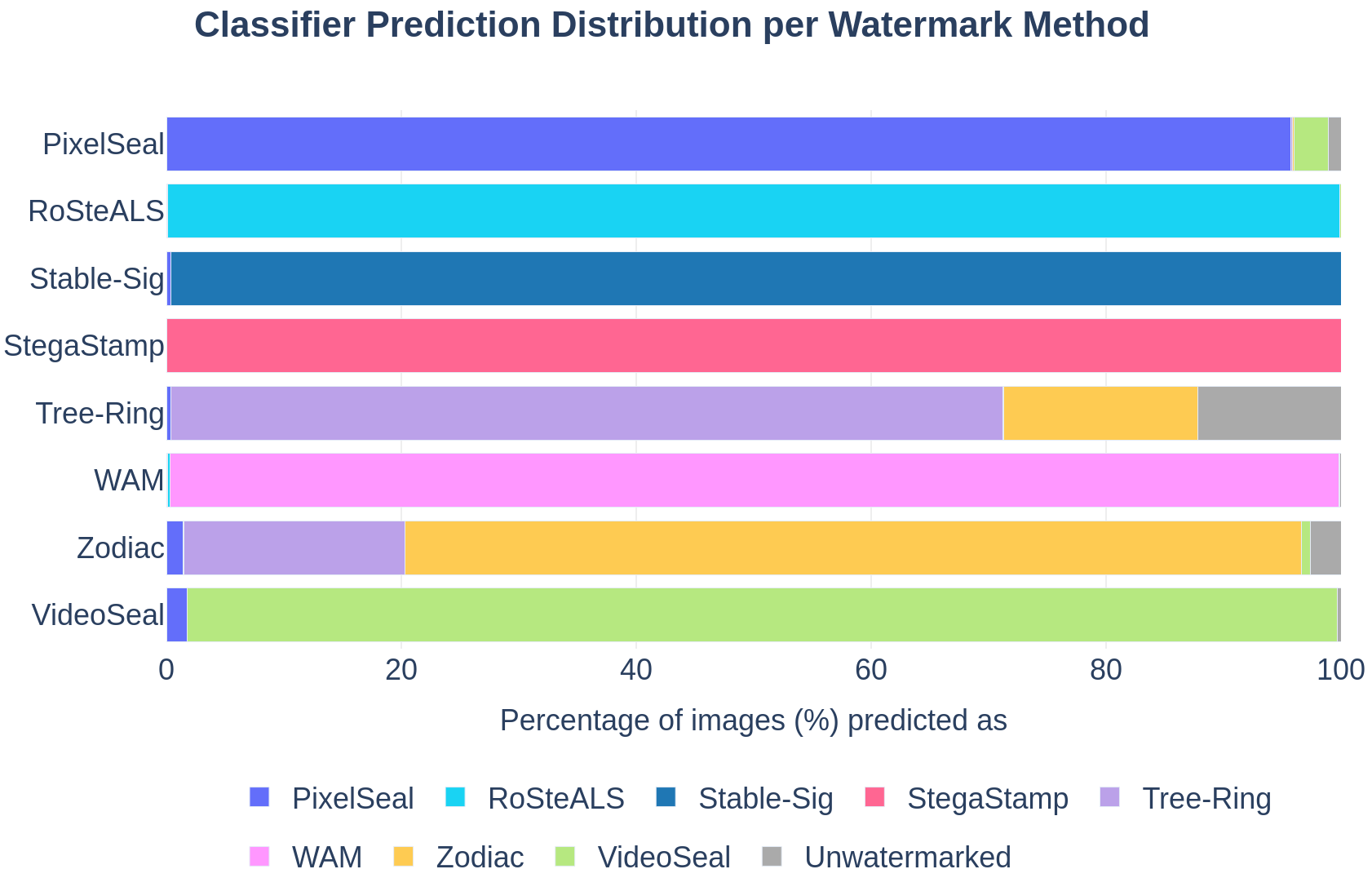}
    \caption{%
        Per-method prediction distribution of the classifier, aggregated over MS-COCO and DiffusionDB.
    }
    \label{fig:cl3_distribution}
\end{figure}

Figure~\ref{fig:cl3_distribution} shows the per-method prediction distribution of the CL3 classifier aggregated over MS-COCO and DiffusionDB. The classifier achieves near-perfect accuracy for Pixel Seal, RoSteALS, Stable Signature, StegaStamp, WAM, and Video Seal, confirming that these methods leave clear scheme-specific fingerprints that are easily recoverable. Notably, despite Pixel Seal and Video Seal sharing a learned encoder-decoder architecture, the classifier reliably distinguishes between them. The only notable confusion occurs between Tree-Ring and Zodiac: approximately 70\% of Tree-Ring images are correctly classified, with the remainder misclassified predominantly as Zodiac, and conversely around 20\% of Zodiac images are misclassified as Tree-Ring. We attribute this mutual confusion to their shared embedding mechanism: both methods inject watermarks into the frequency spectrum of the diffusion latent during sampling and therefore seem to leave similar artifacts. As discussed in Section~\ref{sec:pipeline}, this confusion is operationally benign since both methods are routed to the same removal attack anyway. The only consequential failure mode is misclassification as \emph{Unwatermarked} (gray segment), visible for Tree-Ring, as such images bypass the attack pipeline entirely and their watermark is left intact. 

\subsection{Classifier Training Details and Ablations}
\label{app:classifier_ablations}

\paragraph{Training Details for cl1} We fine-tune ConvNeXt-V2 Large end-to-end using the AdamW optimizer \cite{loshchilov2018decoupled} ($lr=2\times10^{-5}$, weight decay=$0.01$) with a cosine schedule and 10\% linear warmup \cite{loshchilov2017sgdr}. We maintain an effective batch size of 16 through gradient accumulation and utilize mixed-precision (fp16) training throughout. Final regularization included label smoothing of 0.1 and early stopping with a patience of five epochs.

We experiment with two further training setups on top of the baseline presented in \Cref{sec:classify-wm} (referred to as cl1). Both were attempts to improve out-of-distribution behavior without harming the in-distribution split. Neither dominates cl1 across the board, so we keep cl1 as our main classifier and report the other two attempts here for completeness.

\paragraph{cl2: Extra Unwatermarked Images.} For this setup, we retain the architecture, optimizer,
learning rate schedule and the base 500-image-per-class 80/10/10 split as our primary classifier (cl1). However, we add 500
additional unwatermarked DiffusionDB images to the \emph{training} fold. This adjustment aims to broaden the
decision boundary for the ``unwatermarked'' class, which represents the primary failure mode for cl1.

\paragraph{cl3: Augmentation, Partial Freezing, Macro-F1 Model Selection.}
Building upon the cl2 dataset (which includes the extra unwatermarked training images), this setup introduces on-the-fly data augmentations during training. Specifically, we apply random resized crops (scale~$\in[0.7,1.0]$),
horizontal flips, color jitter ($\pm0.1$), JPEG compression ($q\in[60,95]$, 
applied with probability~0.5), Gaussian blur ($\sigma\le 1.0$, applied
with probability~0.2). Furthermore, we freeze the first two stages of the model backbone and select the
best checkpoint based on the macro-F1 score instead of overall accuracy. The aim of this configuration is 
to make the classifier invariant to the kinds of post-processing real images undergo in the wild.

\begin{table}[h]
\centering
\small
\caption{Per-class recall on the three test sets for the three classifier training setups.}
\label{tab:clf_ablations}
\begin{tabular}{lccc|ccc|ccc}
\toprule
& \multicolumn{3}{c|}{DiffusionDB (in-dist.)} & \multicolumn{3}{c|}{MS-COCO (SD~2.1)} & \multicolumn{3}{c}{MS-COCO~(SD~3.5)} \\
Class & cl1 & cl2 & cl3 & cl1 & cl2 & cl3 & cl1 & cl2 & cl3 \\
\midrule
Pixel Seal     & 0.960 & 0.940 & 0.820 & 0.930 & 0.896 & 0.802 & 0.824 & 0.794 & 0.738 \\
RoSteALS      & 1.000 & 1.000 & 1.000 & 0.998 & 0.998 & 1.000 & 0.996 & 0.996 & 1.000 \\
Stable Sig.   & 1.000 & 1.000 & 1.000 & 0.992 & 0.998 & 0.994 & ---   & ---   & ---   \\
StegaStamp    & 1.000 & 1.000 & 1.000 & 1.000 & 1.000 & 0.996 & 1.000 & 1.000 & 0.998 \\
Tree-Ring     & 0.780 & 0.820 & 0.640 & 0.728 & 0.660 & 0.730 & ---   & ---   & ---   \\
WAM           & 1.000 & 1.000 & 0.980 & 0.992 & 1.000 & 0.992 & 0.966 & 0.992 & 0.984 \\
ZoDiac        & 0.980 & 0.920 & 0.900 & 0.634 & 0.536 & 0.242 & 0.490 & 0.338 & 0.196 \\
Video Seal     & 1.000 & 1.000 & 0.980 & 0.968 & 0.978 & 0.912 & 0.986 & 0.976 & 0.888 \\
Unwatermarked & 0.860 & 0.880 & 0.920 & 0.660 & 0.770 & 0.682 & 0.070 & 0.268 & 0.528 \\
\midrule
Overall acc.\ & 0.953 & 0.951 & 0.916 & 0.878 & 0.871 & 0.817 & 0.762 & 0.766 & 0.762 \\
Macro-F1      & 0.953 & 0.951 & 0.915 & 0.876 & 0.864 & 0.797 & 0.579 & 0.602 & 0.761 \\
\bottomrule
\end{tabular}
\end{table}

We observe that classifier \textbf{cl2} (extra real images)
reshuffles the recall budget exactly as expected: unwatermarked
recall improves on every test set (e.g.\ MS-COCO 0.660~$\rightarrow$~0.770, SD~3.5 0.070~$\rightarrow$~0.268), at the cost of small drops on the
hardest watermark classes (ZoDiac and Tree-Ring on MS-COCO). The trade is roughly neutral overall. Classifier \textbf{cl3} (augmentation + partial freezing
+ macro-F1 selection) goes much further in the same direction: it delivers the strongest unwatermarked recall on every test set (particularly under the SD~3.5 shift, 0.070~$\rightarrow$~0.528, and correspondingly the highest SD~3.5 macro-F1 of the three) but it pays for this with substantially weaker recall on Pixel Seal, ZoDiac and Video Seal across the board. In effect, the JPEG/blur/color-jitter augmentation makes the classifier insensitive to fine pixel-level artifacts, which is helpful for the ``no watermark'' decision but actively counter-productive
for identifying watermarks whose signal lives in exactly those fine-grained statistics.

For the threat model in this paper the relevant criterion is recall on the watermarked classes; an unwatermarked image misclassified as a watermark costs only a small quality hit, while a missed watermark sends the attacker down the wrong removal pipeline. By that criterion, cl1 is the
strongest of the three across DiffusionDB and MS-COCO, ties with cl2 on SD~3.5, and we therefore use cl1 throughout the main paper. %

\section{Quality Degradation Caused by Watermarking Methods} \label{app:quality_degradation}

\begin{figure}
    \centering
    \includegraphics[width=\linewidth]{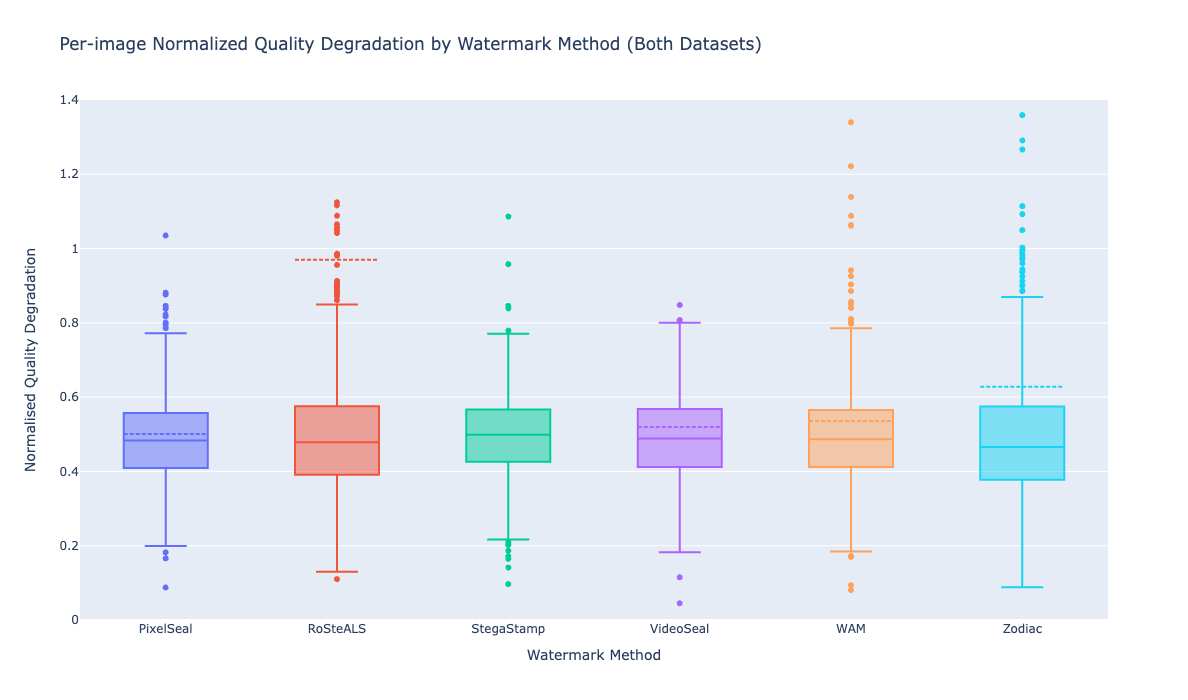}
    \caption{Per-image normalized quality degradation introduced by each watermarking method, aggregated over MS-COCO and DiffusionDB. Each box shows the interquartile range, with the solid line indicating the median and the dashed line the mean. Methods with heavier right tails and wider spreads embed a stronger signal on a subset of images, making them harder to overwrite as victims but potentially more disruptive as attack watermarks.}
    \label{fig:placeholder}
\end{figure}

To gain further insight into our watermark removal results, we examine the per-image normalized quality degradation introduced by each watermarking method (for the set of methods that are acting as both \emph{victim} and \emph{attack} in our experiments) and present the plots in \Cref{fig:placeholder}. The key differences between methods lie in spread and tail behavior rather than the median, which is similar across all six methods. Methods with heavier right tails and wider spreads are likely to be both harder to remove as victims and more disruptive as attackers, whereas tightly distributed methods offer more predictable but potentially more easily targeted behavior. RoSteALS, WAM and ZoDiac exhibit the heaviest right tails and largest IQRs, indicating that on a non-trivial fraction of images, these methods introduce substantially higher distortion (and therefore a stronger embedded signal) than the rest. This may translate to stronger overwriting capability when used as attack watermarks, occasionally at the cost of quality loss, but equally makes them harder to remove as victims. 

This is consistent with our results: Video Seal, StegaStamp, and Pixel Seal, which show the most compact distributions, were successfully overwritten by themselves with bit accuracy reduced to near-chance level and TPR close to zero, while RoSteALS, WAM, and ZoDiac maintained high TPR despite mean bit accuracy dropping from 99\% to around 70\%. The asymmetry between median and mean visible for RoSteALS and ZoDiac, where the mean is pulled above the median by the upper tail, points to a small subset of images carrying a disproportionately strong watermark signal---likely the same images anchoring TPR at the 1\% FPR threshold even when the attack succeeds broadly.

\begin{figure}
    \centering
    \includegraphics[width=\linewidth, trim={0.35cm 0.7cm 0.7cm 0cm}, clip]{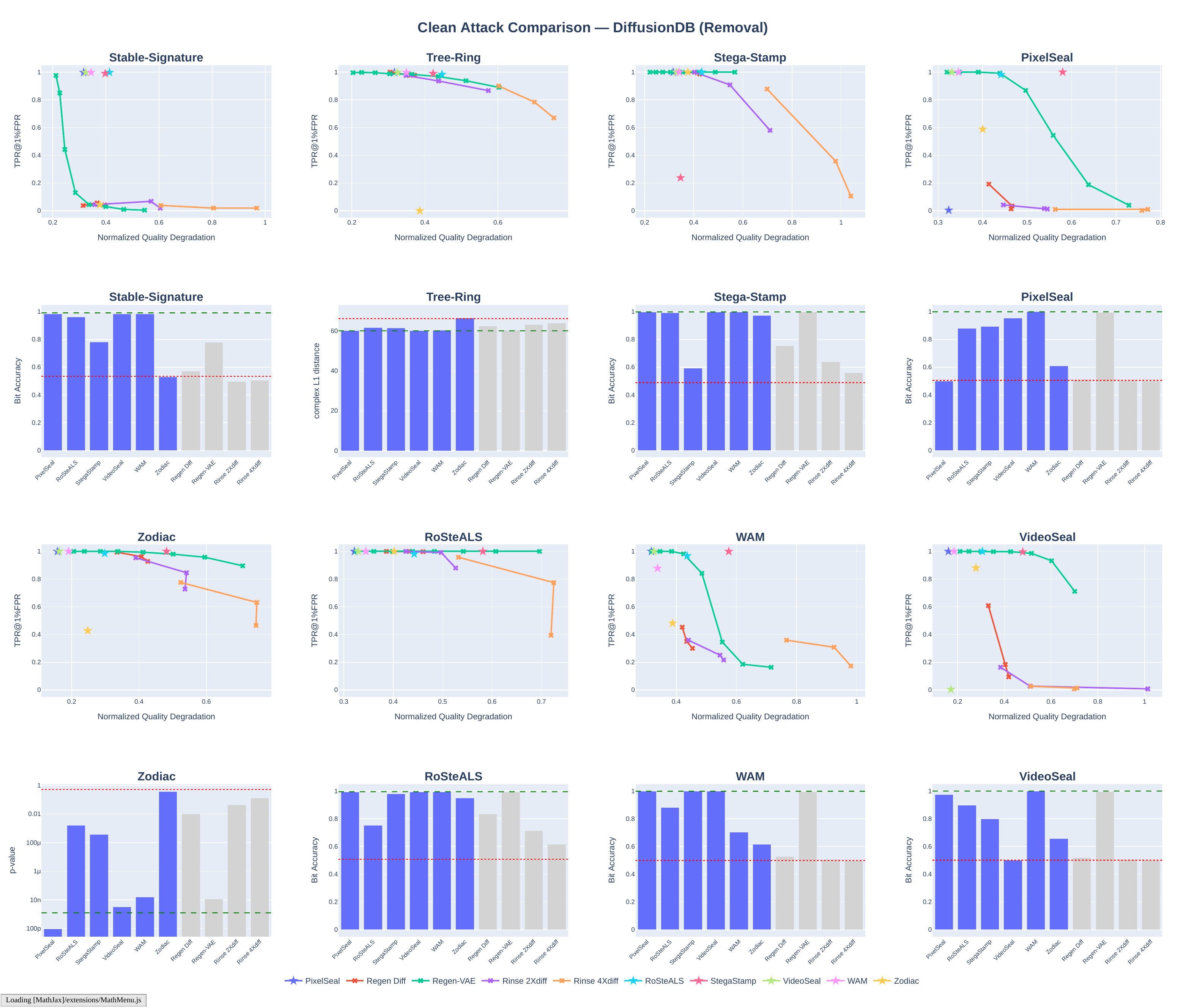}
    \caption{Performance vs.\ image quality degradation on DiffusionDB. Each watermarked image is attacked by layering a second watermark on top, as well as by diffusion-based baseline attacks. Below, we report bit accuracy for message-based methods and $\ell_1$-distance\slash $p$-value for Tree-Ring and ZoDiac, with watermark attacks shown in blue and baseline attacks in gray. DiffusionDB results closely mirror the MS-COCO findings reported in \Cref{fig:waw_big_figure}. For in-processing watermarks, ZoDiac again reduces TPR to near zero for both Tree-Ring and Stable Signature at low perceptual cost. For post-processing watermarks, same-method reapplication remains consistently the most effective attack: StegaStamp, Pixel Seal, and Video Seal are again fully overwritten by a second embedding of themselves, while ZoDiac, RoSteALS and WAM retain the highest robustness. Minor quantitative differences between datasets are expected given the distributional shift (\protect\eg, we observe that ZoDiac worked here slightly better as an attack on WAM), but the overall ranking of attacks and victims is preserved. This cross-dataset consistency supports the general attack policy proposed in \cref{sec:pipeline}: apply ZoDiac against in-processing victims and reapply the identified method against post-processing victims (which is also supported by the results of the end-to-end attack DiffusionDB, see~\Cref{sec:pipeline}).}
    \label{fig:waw_big_figure_diffdb}
\end{figure}

\subsection{Computing Normalized Quality Degradation} \label{app:calculate_nqd}

Following \cite{an2024waves}, we normalize each image quality metric to a common scale before aggregation. Concretely, the 10th and 90th percentiles of each metric's values (computed over all attacked images in our evaluation set) are mapped to 0.1 and 0.9, respectively, with linear interpolation in between. This choice is motivated by the approximately linear behavior of the empirical CDFs between these quantiles, while distributions outside this range are often non-linear (most notably for PSNR). All metrics are oriented so that higher normalized values correspond to greater quality degradation. The normalized scores are then averaged across metrics to produce a single Normalized Quality Degradation score, where lower values indicate less perceptual distortion.

\section{Attacked Examples}

\begin{figure*}[h]
    \centering
    \setlength{\tabcolsep}{2pt}
    \begin{tabular}{ccc}
        \includegraphics[width=0.328\textwidth]{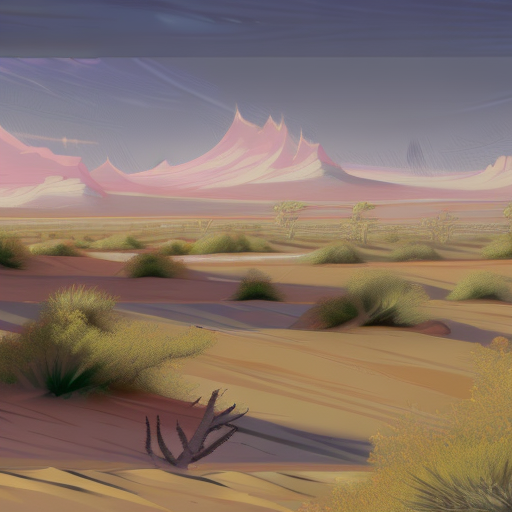} &
        \includegraphics[width=0.328\textwidth]{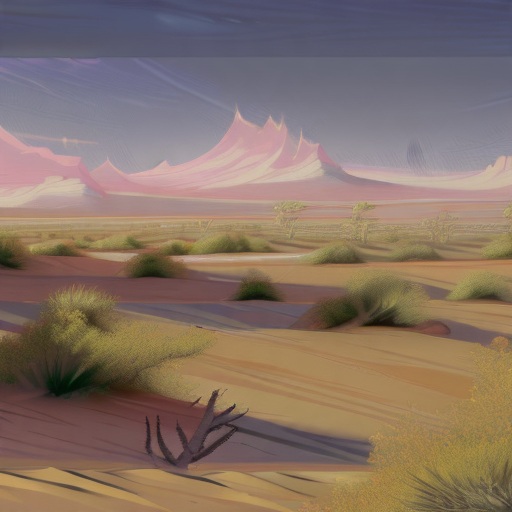} &
        \includegraphics[width=0.328\textwidth]{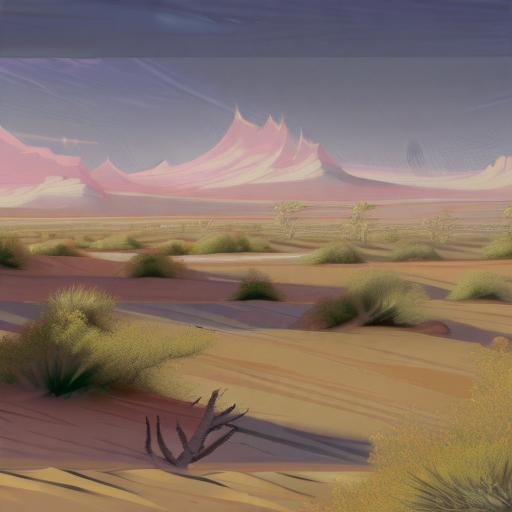} \\
        {\small Unwatermarked image} &
        {\small Watermarked with Pixel Seal} &
        {\small Pixel Seal + Pixel Seal} \\[4pt]
        \includegraphics[width=0.328\textwidth]{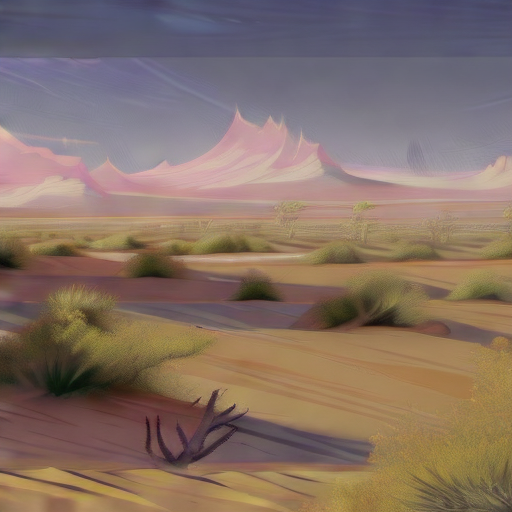} &
        \includegraphics[width=0.328\textwidth]{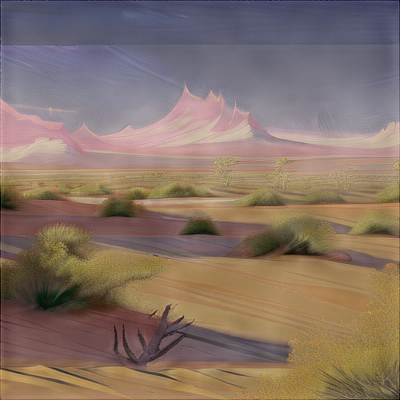} &
        \includegraphics[width=0.328\textwidth]{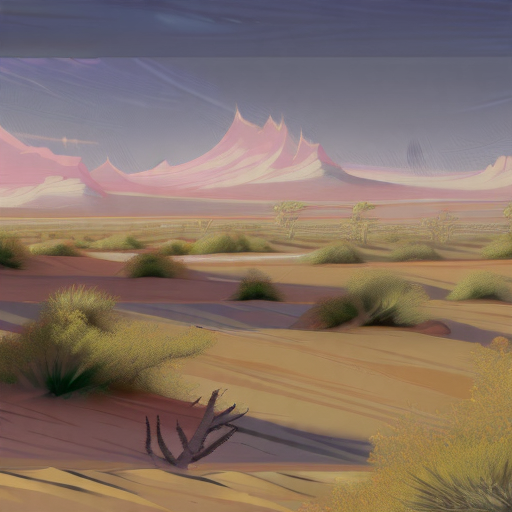} \\
        {\small Pixel Seal + RoSteALS} &
        {\small Pixel Seal + StegaStamp} &
        {\small Pixel Seal + Video Seal} \\[4pt]
        \includegraphics[width=0.328\textwidth]{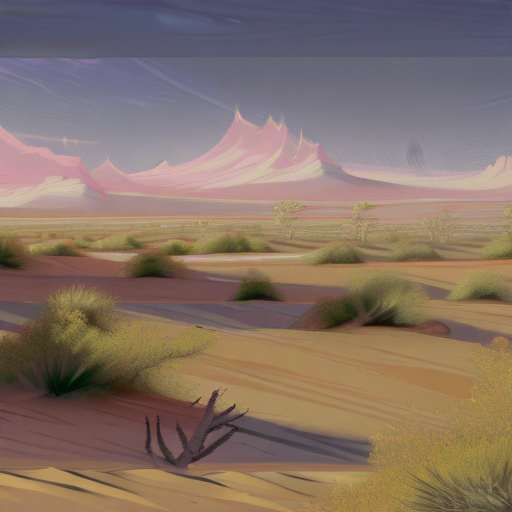} &
        \includegraphics[width=0.328\textwidth]{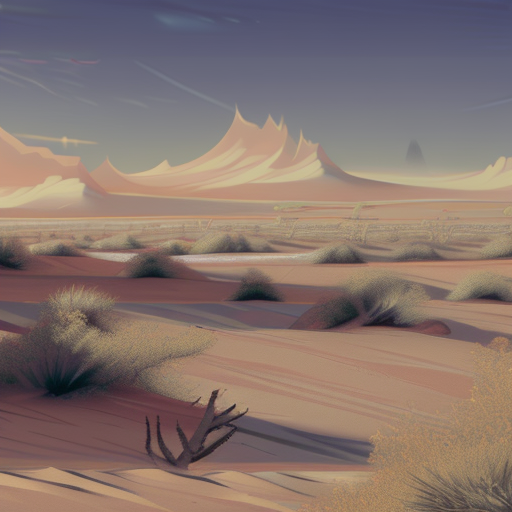} &
        \includegraphics[width=0.328\textwidth]{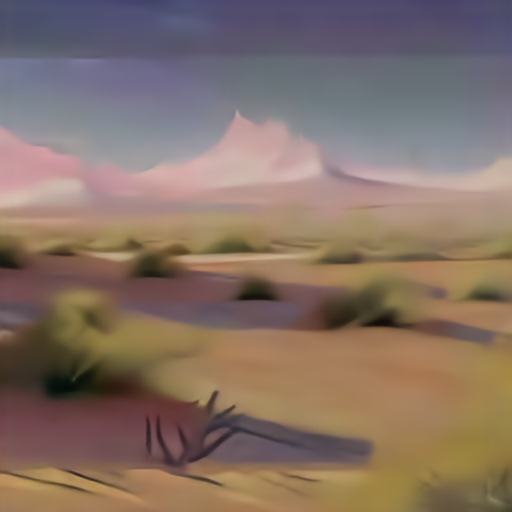} \\
        {\small Pixel Seal + WAM} &
        {\small Pixel Seal + ZoDiac} &
        {\small Pixel Seal + Regen-VAE-1} \\[4pt]
        \includegraphics[width=0.328\textwidth]{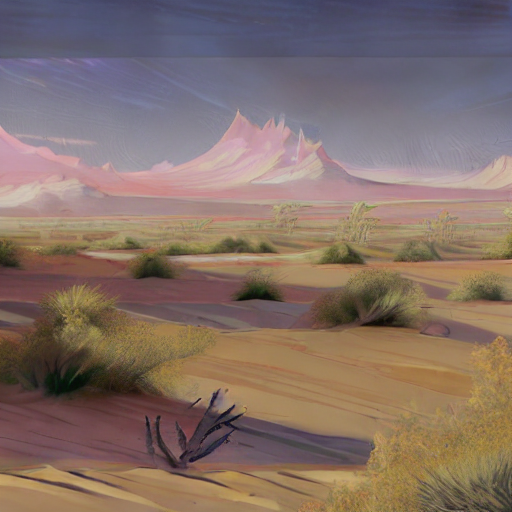} &
        \includegraphics[width=0.328\textwidth]{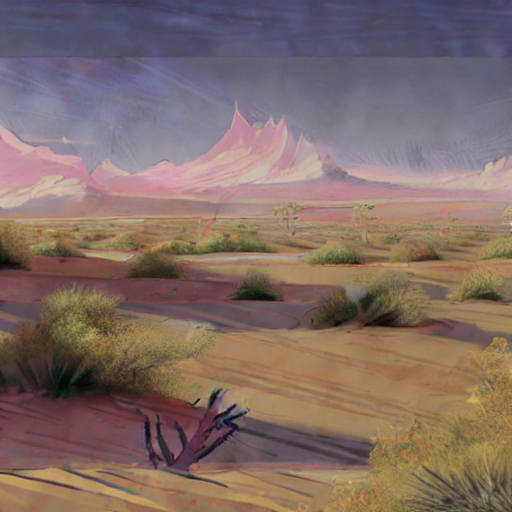} &
        \includegraphics[width=0.328\textwidth]{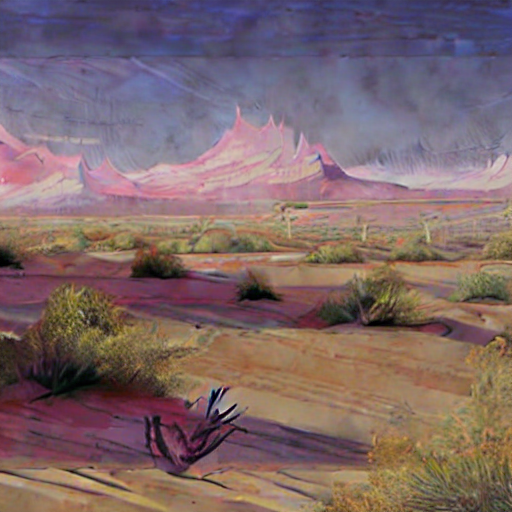} \\
        {\small Pixel Seal + Regen-Diff-100} &
        {\small Pixel Seal + Rinse-2x-Diff-60} &
        {\small Pixel Seal + Rinse-4x-Diff-60} \\
    \end{tabular}
    \caption{Visual examples of all attacks from \Cref{sec:policy} applied to a single DiffusionDB image watermarked with Pixel Seal. Notation \texttt{A + B} denotes an image first watermarked with method~A (\emph{victim watermark}) and subsequently watermarked with method~B (\emph{attack watermark}).}
    \label{fig:examples}
\end{figure*}

In \Cref{fig:examples}, we present visual examples from DiffusionDB illustrating the effect of all attacks on a single image watermarked with Pixel Seal. The first row shows the unwatermarked image, the Pixel Seal-watermarked image, and the re-watermarked image (where re-watermarking means applying Pixel Seal a second time on top of the already watermarked image, but embedding a different message). We observe that across these three versions, no perceptible visual difference is apparent. It is therefore noteworthy that while the original watermark is successfully detected in the second image, it is no longer detectable in the third. The baseline attacks (shown at the perturbation level required to suppress watermark detection) introduce severe visual artifacts that substantially degrade image quality. Although Regen-Diff-100 causes less visible damage than the other baseline attacks, it also fails to reduce the TPR as effectively, in contrast to re-watermarking with Pixel Seal. ZoDiac introduces a subtle shift in color while otherwise leaving the image largely artifact-free. This is because ZoDiac re-synthesizes the image through a guided denoising process, which may alter low-level color statistics. Overall, our re-watermarking attack achieves significantly lower perceptual degradation than the baseline attacks while more effectively suppressing the original watermark.

\section{Computational Resources}
\label{app:compute}

All experiments were run on an internal SLURM-managed HPC cluster. Each job was 
scheduled on a single GPU node, requesting one NVIDIA A100 (80\,GB), H100 (80\,GB), or L40S (48\,GB) GPU together with 4--8 CPU cores and 32--64\,GB of system RAM. Storage requirements amounted to approximately 2\,TB for source images, attacked images, and decoded result JSONs.

Per-experiment runtime varied substantially by attack type. Embedding-based re-watermarking attacks processed 500 images in 2--10 minutes per (attack, strength, victim) combination on a single A100. Diffusion-based regeneration attacks (Regen-Diff, Rinse-2xDiff, Rinse-4xDiff) required 30--90 minutes per such combination, and the optimization-based ZoDiac attack took approximately 2 hours per 100 images ($\approx$10 GPU-hours per dataset--victim pair). Decoding and quality-metric passes (PSNR, SSIM, LPIPS, CLIP-FID, aesthetic and artifact scores, CLIP-score) took 1--20 minutes per job.

\section{Limitations} \label{app:limitations}

While our evaluation covers 48 attack-victim watermark pairs, we acknowledge that additional watermarking schemes and attacks exist beyond the scope of this work. However, our selection spans representative methods across distinct watermarking families, providing broad coverage.
Furthermore, our attack ranking methodology is inherently tied to TPR thresholds and image quality metrics drawn from established literature and experimental practice. Alternative quality metrics (e.g., MSE, Watson-DFT) or different threshold choices may yield different rankings.

\begin{table}[t]
\caption{Decoder metrics on DiffusionDB for our attack vs.\ baselines. BA\textsuperscript{clean}, BA\textsuperscript{unwm}, BA\textsuperscript{atk} denote the bit accuracy on clean watermarked images, unwatermarked images and attacked watermarked images, respectively. TPR is reported at 1\%FPR and \#mis denotes the number of images misclassified as unwatermarked. Bold indicates the strongest attack metric achieved per victim method. Effective attacks yield lower BA and TPR, except for Tree-Ring (\textsuperscript{*}$\ell_1$ distance) and ZoDiac (\textsuperscript{**}$p$-value) where higher is better. Baseline strengths are chosen to not exceed our attack's quality degradation, though Rinse 4$\times$Diff still exceeds it even at the lowest strength.}
\label{tab:pipeline_tab_diffdb}
\centering
\footnotesize
\setlength{\tabcolsep}{4pt}
\resizebox{\textwidth}{!}{%
\begin{tabular}{lcccccp{0.05cm}ccp{0.05cm}ccp{0.05cm}ccp{0.05cm}cc}
\toprule
&&& \multicolumn{3}{c}{\textbf{Our Attack}} & & \multicolumn{2}{c}{\textbf{Regen-Diff 100}} & & \multicolumn{2}{c}{\textbf{RegenVAE 2}} & & \multicolumn{2}{c}{\textbf{Rinse 2$\times$Diff 20}} & & \multicolumn{2}{c}{\textbf{Rinse 4$\times$Diff 20}} \\
\cmidrule(lr){4-6} \cmidrule(lr){8-9} \cmidrule(lr){11-12} \cmidrule(lr){14-15} \cmidrule(lr){17-18}
\textbf{Method} & BA\textsuperscript{clean} & BA\textsuperscript{unwm} & BA\textsuperscript{atk} & TPR & \#mis & & BA\textsuperscript{atk} & TPR & & BA\textsuperscript{atk} & TPR & & BA\textsuperscript{atk} & TPR & & BA\textsuperscript{atk} & TPR \\
\midrule

Pixel Seal  & 0.999 & 0.504 & \textbf{0.503} & \textbf{0.008} & 2  & &  {0.510} &  {0.010} & &  {0.534} &  {0.188} & &  {0.518} &  {0.042} & &  {0.506} &  {0.010} \\

RoSteALS   & 0.995 & 0.504 & \textbf{0.749} &  {0.980} & 0  & &  {0.836} &  {0.998} & &  {0.970} &  {1.000} & &  {0.869} &  {1.000} & &  {0.754} & \textbf{0.958} \\

Stable Sig.  & 0.985 & 0.527 &  {0.532} &  {0.014} & 0  & &  {0.571} &  {0.043} & &  {0.570} & \textbf{0.010} & &  {0.577} &  {0.044} & & \textbf{0.508} &  {0.038} \\

StegaSt    & 0.997 & 0.509 & \textbf{0.581} & \textbf{0.218} & 0  & &  {0.754} &  {0.986}& & {0.996} & {1.000} & & {0.841} & {1.000} & & {0.715} &  {0.878}\\

Tree-Ring   & 53.2\textsuperscript{*} & 66.4\textsuperscript{*} & \textbf{64.4\textsuperscript{*}} & \textbf{0.370} & 60 & &  {62.4\textsuperscript{*}} &  {0.978} &&  {62.2\textsuperscript{*}}  &  {0.938}& &  {62.2\textsuperscript{*}} &  {0.976} &&  {63.1\textsuperscript{*}} &  {0.900} \\

Video Seal  & 0.999 & 0.500 & \textbf{0.503} & \textbf{0.008} & 0  & &  {0.521} &  {0.095} & &  {0.680}& {0.932} & &  {0.554}& {0.164} & & {0.512} &  {0.027}\\

WAM        & 0.998 & 0.500 &  {0.693} &  {0.874} & 1  & &  {0.528}&  {0.301}& &\textbf{0.500} & \textbf{0.186} & & {0.550} &  {0.360}& & {0.519} &  {0.360}\\

ZoDiac     & $1.5\text{e}{-15}^{**}$ & $5.1\text{e}{-1}^{**}$  & $\mathbf{5.3\text{e}{-1}}^{**}$ & \textbf{0.294} & 8 && 
 {1.0e-2\textsuperscript{**}} &  {0.928} & &  {5.6e-3\textsuperscript{**}}&  {0.958} & &  {8.3e-3\textsuperscript{**}}&  {0.954}& &  {3.9e-2\textsuperscript{**}}&  {0.776}\\
\bottomrule
\end{tabular}}
\end{table}

\end{document}